\documentclass[aps, prd, twocolumn,showkeys, showpacs,amsmath,amssymb,superscriptaddress]{revtex4-1}
\usepackage{graphicx}
\usepackage{dcolumn}
\usepackage{bm}
\usepackage{epsfig}
\usepackage{subfigure}
\usepackage{color}
\usepackage{epstopdf}

\begin{document}
\title{Lattice Boltzmann model for resistive relativistic magnetohydrodynamics}

\author{F. Mohseni} \email{mohsenif@ethz.ch} \affiliation{ ETH
  Z\"urich, Computational Physics for Engineering Materials, Institute
  for Building Materials, Wolfgang-Pauli-Strasse 27, HIT, CH-8093 Z\"urich
  (Switzerland)}

\author{M. Mendoza} \email{mmendoza@ethz.ch} \affiliation{  ETH
  Z\"urich, Computational Physics for Engineering Materials, Institute
  for Building Materials, Wolfgang-Pauli-Strasse 27, HIT, CH-8093 Z\"urich
  (Switzerland)}

\author{S. Succi} \email{succi@iac.cnr.it} \affiliation{Istituto per
  le Applicazioni del Calcolo C.N.R., Via dei Taurini, 19 00185, Rome
  (Italy),\\and Freiburg Institute for Advanced Studies,
  Albertstrasse, 19, D-79104, Freiburg, (Germany)}

\author{H. J. Herrmann}\email{hjherrmann@ethz.ch} \affiliation{  ETH
  Z\"urich, Computational Physics for Engineering Materials, Institute
  for Building Materials, Wolfgang-Pauli-Strasse 27, HIT, CH-8093 Z\"urich
  (Switzerland)} \affiliation{Departamento de F\'isica, Universidade
  Federal do Cear\'a, Campus do Pici, 60455-760 Fortaleza, Cear\'a,
  (Brazil)}

\date{\today}
\begin{abstract}
 In this paper, we develop a lattice Boltzmann model for relativistic magnetohydrodynamics (MHD). Even though the model is derived for resistive MHD, it is shown that it is numerically robust even in the high conductivity (ideal MHD) limit. In order to validate the numerical method, test simulations are carried out for both ideal and resistive limits, namely the propagation of Alfv\'en waves in the ideal MHD and the evolution of current sheets in the resistive regime, where very good agreement is observed comparing to the analytical results. Additionally, two-dimensional magnetic reconnection driven by Kelvin-Helmholtz instability is studied and the effects of different parameters on the reconnection rate are investigated. It is shown that the density ratio has negligible effect on the magnetic reconnection rate, while an increase in shear velocity decreases the reconnection rate. Additionally, it is found that the reconnection rate is proportional to $\sigma^{-\frac{1}{2}}$, $\sigma$ being the conductivity, which is in agreement with the scaling law of the Sweet-Parker model. Finally, the numerical model is used to study the magnetic reconnection in a stellar flare. Three-dimensional simulation suggests that the reconnection between the background and flux rope magnetic lines in a stellar flare can take place as a result of a shear velocity in the photosphere.
 
\end{abstract}


\maketitle
\section{Introduction}
\label{Introduction}

Magnetic fields are an essential component of many astrophysical phenomena, such as relativistic jets \cite{doi:10.1146/annurev.astro.37.1.409}, active galactic nuclei \cite{antonucci1993unified}, gamma ray bursts \cite{RevModPhys.76.1143}, pulsar winds \cite{Camus11122009}, and stellar flares \cite{masada2010solar}. Since in most of these phenomena the plasma is electrically neutral and the characteristic times between collisions are much smaller than the typical time scale of the system, the magnetohydrodynamics (MHD) approximation is appropriate. Due to the fact that relativistic effects play a major role in the dynamics of these phenomena, relativistic MHD description is of special interest in this perspective. Except for some simple geometries, most of the studies are based on numerical simulations, since the equations of relativistic MHD are extremely difficult to solve analytically. 

Ideal MHD is defined as the limit where the electrical conductivity $\sigma$ goes to infinity (electrical resistivity $\eta \equiv 1/\sigma \rightarrow 0$). In this framework many numerical models have been developed over the last decade dealing with the ideal relativistic MHD \cite{komissarov1999numerical,duez2005relativistic,farris2008relativistic}. As it will be explained later, the ideal MHD assumption not only makes the solution of the relativistic MHD considerably simpler, but it is also a fairly good approximation for many high-energy phenomena. However, in several situations such as neutron star mergers \cite {faber2012binary} or central engines of gamma ray bursts \cite{macfadyen1999collapsars}, the conductivity can be small and the ideal MHD assumption is not valid any longer. 

More importantly, magnetic reconnection only takes place when resistivity exists in the plasma. This process is the driver of explosive events in astrophysical plasmas, in which magnetic field lines break and reconnect and  the magnetic field topology goes through a sudden change. During this process, plasma releases the magnetic energy and converts it into thermal and kinetic energy on a short timescale. Magnetic reconnection has been proposed to have an influential role in many astrophysical observations, namely as a cause of particle acceleration in extragalactic jets \cite{jaroschek2004relativistic}, as a source of high energy emission \cite{KirkAA}, as an explanation of the rapid variability observed in active galactic nuclei \cite {DrenkhahnAA} and many others. Therefore, studying magnetic reconnection is of great importance, especially considering the fact that the relativistic theory of magnetic reconnection is not yet well established and its mechanism is poorly understood \cite{zenitani2009two}. It should be mentioned that, numerical results of ideal relativistic MHD models sometimes show magnetic reconnection, which is non-physical, since it is caused by numerical resistivity and hence depends on the details of the numerical scheme and resolution \cite{komissarov2007multidimensional}. 

Therefore, there is a strong interest in developing numerical models for resistive relativistic MHD. However, the corresponding governing equations turn out to be numerically very challenging, since the source terms in the equations become stiff, especially when the conductivity is not small \cite{komissarov2007multidimensional}. This is the natural consequence of the fact that the time-scale of the diffusive effects and the overall dynamical time-scale are of the same order. Thus, it is not surprising that the first numerical models for resistive relativistic MHD appeared only in 2006 \cite{watanabe2006two} and 2007 \cite{komissarov2007multidimensional}. In the latter, the fluxes are computed by using the Harten-Lax-van-Leer (HLL) approximate Riemann solver and Strang's splitting technique is used for the stiff source
terms. Later on, a numerical method which uses an implicit-explicit (IMEX) Runge-Kutta method to solve the stiff source terms in the equations is proposed in Ref.\cite{palenzuela2009beyond}. Also, a unified framework for the construction of one-step finite volume and discontinuous Galerkin schemes for the resistive relativistic MHD is introduced in Ref.\cite{dumbser2009very}. More recently, a different approach has been suggested in Ref.\cite{takamoto2011new}, where the method of characteristics is used to solve the Maxwell equations. Additionally the role of the equation of state in the resistive relativistic MHD is investigated in Ref.\cite{mizuno2013role}.

All the above mentioned models are based on solving the macroscopic governing equations of the resistive relativistic MHD. However, in the last few years, new approaches based on lattice Boltzmann (LB) methods have been developed to study relativistic hydrodynamics \cite{mendoza2010fast,mendoza2013relativistic,mohseni2013lattice}. Like the other LB methods, they are based on a minimal lattice version of the Boltzmann kinetic equation, where representative particles stream and collide on the nodes of a regular lattice with sufficient symmetry to reproduce the correct macroscopic equations. The advantages of these models compared to the conventional methods are their mathematical simplicity, computational efficiency on parallel computers, and easy handling of complex geometries.

In this paper, we develop a LB model for relativistic MHD. The model is proposed for resistive MHD but, as we will show later, it is robust enough in the ideal MHD limit as well. The hydrodynamic part is based on the model proposed in Ref.\cite{mohseni2013lattice}, with several extensions, namely to include the contribution of electromagnetic fields in the energy-momentum tensor (corresponding to adding the Lorentz force and Joule heating in the macroscopic equations) as well as to deal with a more general equation of state. For the electromagnetic part, i.e., solving the Maxwell equations, the LB model for electrodynamics proposed in Ref.\cite{mendoza2010three} is modified and extended for coupling with the fluid equations and to include the relativistic Ohm's law. It should be mentioned that in the non-relativistic context, there are several LB models for resistive MHD \cite{succi1991lattice, dellar2002lattice, breyiannis2004lattice}, especially for simulating magnetic reconnection \cite{chen1991lattice,martinez1994lattice}. Our goal is to bring the well-known advantages of the lattice Boltzmann schemes to the context of resistive relativistic MHD. 

The model is validated using numerical tests for the ideal MHD and resistive MHD regimes. In particular, the propagation of Alfv\'en waves in high conductivity media, and the evolution of current sheets in resistive media are validated against analytical solutions. Moreover, as an application for the model, the magnetic reconnection process driven by Kelvin-Helmholtz (KH) instability is studied in detail. The KH instability is one of the fundamental hydrodynamic instabilities which occurs during the shear flow of a uniform fluid, or two fluids with different densities. It was discovered independently by Kelvin and Helmholtz in the 19th century \cite{Kelvinbook,professor1868xliii}. It is  believed that the KH instability appears in the solar-wind interaction with the Earth's magnetosphere which can influence the magnetic reconnection process that takes place there \cite{faganello2012magnetic}. Moreover, the KH instability has been widely investigated for astrophysical applications e.g., astrophysical jet morphology \cite{lobanov2001cosmic} motion of interstellar clouds \cite{vietri1997survival} and clumping in supernova remnants \cite{wang2001instabilities}, where in many of these phenomena, relativistic and magnetic field effects cannot be ignored. Here, we study the KH instability as a potential driver of magnetic reconnection. In the non-relativistic context this has been discussed in Refs. \cite{keppens1999growth,nakamura2006magnetic}. Here we focus on this phenomenon in the relativistic context and we are interested in the effects of the hydrodynamics parameters, i.e., shear velocity and density ratio, as well as the effects of the conductivity on the magnetic reconnection rate. 

Furthermore, the results of a three-dimensional simulation of the magnetic reconnection in a stellar flare driven by a shear velocity on its photosphere are presented. It has been suggested that solar flares are good prototypes for stellar flares in relativistic stars like neutron stars \cite{masada2010solar}. Therefore, a solar type initial condition, consisting of a potential quadrupole background field and a flux rope \cite{takahashi2011scaling}, is chosen to study the stellar flare. We show that the shear velocity on the photosphere of the star can cause the magnetic reconnection to take place between the flux rope and the background magnetic field lines.  

The paper is organised as follows: in Sec.\ref{The resitive relativistic MHD equations}, the basic equations for resistive relativistic MHD are presented; in Sec.\ref{Lattice Boltzmann model for resistive relativistic MHD}, the development of a lattice Boltzmann model for solving the governing equations is elaborated; in Sec.\ref{Test simulations and application}, validation tests and the aforementioned applications of the model are presented; and finally in Sec.\ref{Conclusions}, as a conclusion, an overall discussion of the model and the results are provided.

\section{The resistive relativistic MHD equations}
\label{The resitive relativistic MHD equations}
The equations of motion for resistive relativistic MHD can be written in the covariant form as
\begin{equation}\label{Density_cons}
\partial_\mu N^\mu=0,
\end{equation}
where $N^\mu=n U^\mu$ is the density current with $n$ the mass density, $(U^\mu) = (c, \vec{u}) \gamma (u)$ the four-velocity, $\vec{u}$ the three-dimensional velocity,
$c$ the speed of light, $\gamma(u)=1/\sqrt{1-u^2/c^2}$ the Lorentz's factor, and  
\begin{equation}\label{Energy_momentum_cons}
\partial_\mu T^{\mu \nu}=0,
\end{equation}
where $T^{\mu \nu}$ is the total energy-momentum tensor defined as the sum of fluid energy-momentum tensor and the contribution of the electromagnetic fields, i.e., 
\begin{equation}\label{El_total}
T^{\mu \nu}= T_{\rm Fluid}^{\mu \nu}+T_{\rm EM}^{\mu \nu},
\end{equation}
with 
\begin{equation}\label{EM_Fluid}
T_{\rm Fluid}^{\mu \nu}=(\epsilon + p) \frac{U^\mu U^ \nu}{c^2}-p \eta^{\mu \nu}+ \pi^{\mu \nu}, 
\end{equation}
where $\epsilon$ is the energy density (including rest mass energy), $p$ is the hydrostatic pressure and $\pi^{\mu \nu}$ is the dissipation tensor. On the other hand 
\begin{equation}\label{EM_EM}
T_{\rm EM}^{\mu \nu}= \epsilon_0 (F^{\mu \rho} F^{\nu}_{\rho}+\frac{1}{4} F^{\rho \sigma}F_{\rho \sigma} \eta^{\mu \nu}),
\end{equation}
where $F^{\mu \nu}$ is the Maxwell electromagnetic tensor defined as
\begin{equation}\label{Maxwell_tensor}
(F^{\mu \nu})=
 \left( \begin{array}{cccc}
0 & -E^x & -E^y & -E^z\\
E^x & 0 & -c B^z & c B^y\\
E^y & c B^z & 0 & -c B^x\\
E^z & -c B^y & c B^x & 0 \end{array} \right) ,
\end{equation}
where $E^i$ and $B^i$ are the electric and magnetic fields, respectively, and $\epsilon_0$ is the permittivity of free space which relates to the permeability of the free space, $\mu_0$, through the relation $c^2 \epsilon_0 \mu_0 = 1$. Note that, throughout this paper, Latin superscripts (subscripts) run over the spatial coordinates, while Greek superscripts (subscripts) run over the four-dimensional (4D) space-time coordinates.

In addition to the mentioned hydrodynamics conservation equations, the governing equations for the electromagnetic fields, i.e., Maxwell equations, also need to be considered, which in the covariant form read as
\begin{equation} \label{Max_cov_Max}
\partial_\nu F^{\mu \nu}=-\mu_0 c I^\mu,
\end{equation}
and 
\begin{equation} \label{Max_cova_Far}
\partial_\nu F^{*\mu \nu}=0,
\end{equation}
where $(I^\mu)=(c \rho_c, \vec{J})$ is the four-vector of electric current with $\rho_c$ the charge density and $\vec{J}$ the three-dimensional electrical current while $F^{*\mu \nu}$ is the Faraday tensor defined as
\begin{equation}
F^{*\mu \nu}=\frac{1}{2}\epsilon^{\mu \nu \lambda \kappa} F_{\lambda \kappa},
\end{equation}
with $\epsilon^{\mu \nu \lambda \kappa}$ the Levi-Civita tensor.

By choosing an appropriate decomposition of the Maxwell tensor, one can show that Eqs.\eqref{Max_cov_Max} and \eqref{Max_cova_Far} yield the familiar Maxwell equations \cite{cercignani}
\begin{equation}\label{Gauss_E}
\vec{\nabla} \cdot \vec{E}=\frac{1}{\epsilon_0}\rho_c,
\end{equation}
\begin{equation}\label{Gauss_B}
\vec{\nabla} \cdot \vec{B} =0,
\end{equation}
\begin{equation}\label{Ampere}
\frac{1}{c^2} \partial_t \vec{E}-\vec{\nabla} \times \vec{B}=-\mu_0 \vec{J}, 
\end{equation}
\begin{equation}\label{Faraday}
\partial_t \vec{B}+ \vec{\nabla} \times \vec{E}=0,
\end{equation}
where Eq.~\eqref{Ampere} is the Maxwell-Ampere and  Eq.~\eqref{Faraday} is the Maxwell-Faraday equation.
The equation for conservation of current
\begin{equation}\label{Current_cons}
\partial_t \rho_c+\vec{\nabla} \cdot \vec{J}=0,
\end{equation}
can be obtained by taking the divergence of Eq.\eqref{Ampere} by considering Eq.\eqref{Gauss_E} and Eq.\eqref{Gauss_B} as constraints.

Furthermore, the coupling between the fluid equations and Maxwell equations is expressed by Ohm's law. In general, the explicit form of the current four-vector $I^\mu$ depends on the properties of the electromagnetic fields as well as the fluid variables. Here we use Ohm's law for a resistive isotropic plasma as \cite{komissarov2007multidimensional}
\begin{equation}\label{Ohm}
\vec{J}=\sigma \gamma \left[ \vec{E}+\vec{u}\times \vec{B}-\frac{(\vec{E}\cdot \vec{u})\vec{u}}{c^2}\right] +\rho_c\vec{u},
\end{equation}
with $\sigma$ being the conductivity of the plasma. It is worth mentioning that in the fluid rest frame Ohm's law becomes
\begin{equation}
\vec{J}=\sigma \vec{E},
\end{equation}
and in the limit of $\sigma \rightarrow \infty$ one can obtain the well-known result for ideal MHD
\begin{equation}\label{Ohm_ideal}
\vec{E}=-\vec{u} \times \vec{B}.
\end{equation}
The major difference between the numerical models for ideal and resistive MHD originates from the fact that, in the ideal case, one can substitute the electric field $\vec{E}$ in all the equations, using a simple algebraic relation, i.e., Eq.\eqref{Ohm_ideal}, and thus one can define the electromagnetic induction four-vector $F^{*\mu\nu}U_\nu$ \cite{qamar2005high}. This leads to a considerably simpler and less expensive numerical algorithm, compared with the resistive MHD.

To summarize the governing equations, and by substituting Eq.\eqref{Ohm} into Eqs.\eqref{Ampere} and \eqref{Current_cons}, we have $12$ equations, i.e., Eqs.\eqref{Density_cons}, \eqref{Energy_momentum_cons}, \eqref{Ampere}, \eqref{Faraday} and \eqref{Current_cons} and $13$ unknowns, i.e., $\vec{u}$, $\vec{B}$, $\vec{E}$,  $\epsilon$, $p$, $n$ and $\rho_c$. This system of equations will be complete by including the equation of state. Here, we consider the ideal gas equation of state \cite{ryu2006equation}
\begin{equation}\label{EOS}
p=(\Gamma-1)(\epsilon-n c^2),
\end{equation}
where, $\Gamma$ is the adiabatic index which for ultrarelativistic temperatures takes the value $4/3$, and for non-relativistic temperatures has the value $5/3$.

\section{Lattice Boltzmann model for resistive relativistic MHD}
\label{Lattice Boltzmann model for resistive relativistic MHD}
In this section we describe our lattice Boltzmann model to solve the aforementioned governing equations.

\subsection*{Relativistic fluid equations}

We start with the description of our LB model to solve the equations of motion of the fluid, i.e., Eqs.\eqref{Density_cons} and \eqref{Energy_momentum_cons}. The relativistic Boltzmann equation, based on the Anderson-Witting collision operator \cite{Anderson1974466} is
\begin{equation}\label{Boltzmann}
  p^\mu \partial_\mu f = -\frac{U_\mu p^\mu}{ \tau c^2}( f - f^{\rm eq} ),
\end{equation}
with the Maxwell-J\"uttner
equilibrium distribution function as
\begin{equation}\label{MJ}
  f^{\rm eq} = A \exp(-p_\mu U^\mu/k_B T),
\end{equation}
where $(p^\mu) = (E/c, \vec{p})$ the
four-momentum, with the energy of the particles $E$ defined by
\begin{equation}
  E=cp^0=\frac{mc^2}{\sqrt{1-u^2/c^2}},
\end{equation}
where $m$ is the rest mass and $u^2$ is the magnitude of the three-dimensional velocity. Here, $f=f(\vec{x},\vec{p},t)$ is the probability distribution function of particles at time $t$ in location $\vec{x}$ and momentum $\vec{p}$, which can be used to calculate both the density current and total energy momentum tensor (see below). Note that in LB models, which are based on a lattice version of the Boltzmann equation, the aforementioned particles are collective degrees of freedom (quasi-particles) which can eventually propagate faster than light, as long the physical flow speed remains subluminal. Also, $\tau$ is the
single relaxation time, $A$ is a normalization constant, $k_B$ is the Boltzmann constant and $T$ is the temperature.

As mentioned, our description to solve the fluid equations is based on the model recently proposed for relativistic LB \cite{mohseni2013lattice}. Several modification and extensions are implemented in order to firstly, use ideal gas equation of state (by using Anderson-Witting as the collision operator and extending the distribution function) and secondly, to include the contribution of electromagnetic fields in the total energy-momentum tensor (by adding corresponding terms to the distribution function). Following Ref.\cite{mohseni2013lattice}, we use changes of variables which read as:
\begin{equation}
  \xi^\mu = \frac{p^\mu/m}{c_s} , \quad \chi^\mu = \frac{U^\mu}{c_s},
\end{equation}
\begin{equation}
  c_s = \sqrt{\frac{k_B T}{m}} , \quad \nu = c/c_s.
\end{equation}

\begin{figure}
\centering 
\includegraphics[width=0.7\columnwidth]{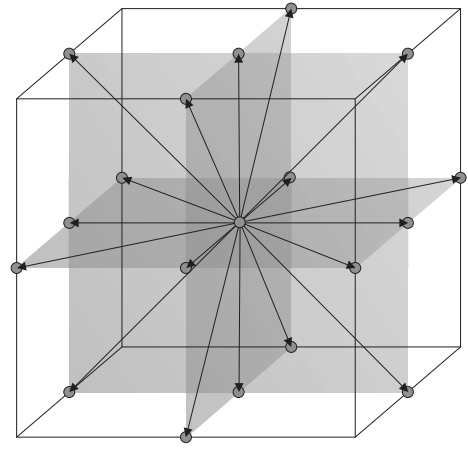} 
\caption{The D3Q19 lattice configuration for the LB model to solve the relativistic fluid equations.}
\label{D3Q19}
\end{figure}

In order to discretize the Boltzmann equation, Eq.\eqref{Boltzmann}, we write $(\xi^{\alpha})=(c_t/c_0, \vec{c}_a) $,
where $c_t$, $c_0$ and $c_a$ are constants related to the size of the
lattice. We use a lattice configuration D3Q19 (19 discrete
vectors in 3 spatial dimensions), which can be expressed as
\begin{equation} \label{cellconfigs1}
  \vec{c}_a=\left \{ \begin{array}{ll}
      (0,0,0) &  i=0; \\
      c_a(\pm 1,0,0)_{FS} &  1\leq i \leq 6; \\
      c_a(\pm 1,\pm 1,0)_{FS} &  7\leq i \leq 18, \end{array} \right.
\end{equation}
where the subscript $FS$ denotes a fully symmetric set of points (see Fig.\ref{D3Q19}).

Using the quadrature rule and knowing that $\sum_i w_i=1$ the discretized weights can be calculated as (see Ref.~\cite{mohseni2013lattice} for the derivation)
\begin{equation}
w_0 = 1+\frac{4 c_t^2 \nu^2}{361 c_0^2}-\frac{c_t^2}{c_a^2 c_0^2} ,
\end{equation}
\begin{equation}
w_i = \frac{c_t^2}{2166 c_0^2 c_a^2} \left(361-8 c_a^2 \nu^2\right) ,
\end{equation}
for $1\leq i \leq 6$, and
\begin{equation}
w_i = \frac{c_t^2 \nu^2}{1083 c_0^2},
\end{equation}
for $7\leq i \leq 18$.  

The aforementioned constants can be calculated as
\begin{equation}
c_a=\frac{\sqrt{19}}{\nu}, \quad c_t/c_0=\frac{\sqrt{27}}{\nu}, \quad c_0=\frac{3}{8}(9-2\sqrt{3}).
\end{equation}

In order to increase the numerical stability of the LB model (particularly for higher velocities) a flux limiter scheme (min mod scheme) is used to discretize the spatial derivative in the streaming term in the Boltzmann equation, i.e. $p^a_i \partial _a f_i$. The min mod scheme efficiently reduces the instability, especially in the presence of discontinuities or large gradients. We have
\begin{equation}
\partial _a (p^a_i f_i) = \frac{1}{|\delta x \vec{e}_a|} \left[ h_i^a(\vec{x}+\delta x  \vec{e}_a) - h_i^a(\vec{x}) \right],
\end{equation}
where $\vec{e}_a$ is a unit vector in the direction of the
corresponding spatial coordinate. For the definition of $h_i^a(\vec{x})$ and more details see Ref.\cite{mohseni2013lattice}.

Therefore, the discretized form of the relativistic Boltzmann equation 
takes the following expression:
\begin{equation}\label{eq:rlb}
\begin{aligned}
f_i(\vec{x}, t+\delta t) &- f_i(\vec{x}, t) + \frac{c_0}{c_t} \frac{\delta t}{\delta x} \left[ h_i^a(\vec{x}+\delta x  \vec{e}_a) - h_i^a(\vec{x}) \right]  = \\ 
&-\frac{c_0 \nu \delta t (\xi_\mu \chi^\mu/\nu^2)}{\tau c_t}[f_i(\vec{x},t+\delta t) - f_i^{\rm eq}(\vec{x},t)] \\ 
& +\frac{c_o \nu \delta t}{c_t} \lambda_i \sum_i \partial_a^2 f_i(\vec{x},t),
\end{aligned}
\end{equation}
where the last term is the bulk viscosity term with
\begin{equation}
 \lambda_i = \left\{ \begin{array}{ll}
0 & i=0 ; \\
 \alpha \delta x & i\not= 0,\end{array} \right. 
\end{equation}
where $\alpha$ is a small constant. The exact value of $\alpha$ for each simulation is reported in Sec.\ref{Test simulations and application}. In fact, this small value of bulk viscosity is sufficient to stabilize the numerical procedure. A central finite difference scheme is used to calculate the second order derivative.  

One can notice that to solve the discretized Boltzmann equation, Eq.\eqref{eq:rlb}, the discretized equilibrium distribution function is required. The discretized equilibrium distribution function to recover $T^{\mu \nu}_{Fluid}$ has been proposed in Ref.\cite{mohseni2013lattice}. To include the electromagnetic contribution to the energy-momentum tensor, i.e., $T^{\mu \nu}_{EM}$, additional terms need to be added to the discretized distribution function. Let us elaborate the contribution of electromagnetic fields in the energy-momentum tensor by providing the components of $T^{\mu \nu}_{EM}$ using Eqs. \eqref{EM_EM} and \eqref{Maxwell_tensor}. Thus, we have
\begin{equation}
T^{00}_{EM}=\frac{\epsilon_0}{2}(E^2+c^2 B^2),
\end{equation}
\begin{equation}
T^{0i}_{EM}=\frac{1}{\mu_0 c}(\vec{E}\times\vec{B})^i,
\end{equation}
\begin{equation}
T^{ij}_{EM}=\epsilon_0 \left[ -E^i E^j-c^2 B^i B^j+\frac{1}{2}(E^2+c^2 B^2)\delta^{ij} \right],
\end{equation}
where $E^2$ and $B^2$ are the magnitude of the electric and magnetic fields, respectively, and $\delta^{ij}$ is the Kronecker delta. Note that, adding these terms to the total energy-momentum tensor corresponds to adding the Lorentz force and Joule heating to the macroscopic equations. The following expression shows the complete discretized equilibrium distribution function which recovers the total energy-momentum tensor with the ideal gas equation of state 
\begin{equation}\label{Dis_fun_EM}
\begin{aligned}
  f^{\rm eq}_i &= \frac{3}{4}(\epsilon+p)\frac{c_0^2}{c_t^2} w_i\bigg\{ 1+ \frac{3(\Gamma-1)(\epsilon-n c^2)-\epsilon}{(\Gamma-1)(\epsilon-n c^2)+\epsilon} \\
  &+ \frac{361 [\epsilon-(\Gamma-1)(\epsilon-n c^2)]}{33[(\Gamma-1)(\epsilon-n c^2)-\epsilon]} \delta_{i0}+ c_a^x c_a^y\chi^x\chi^y \\
  &+c_a^x c_a^z\chi^x\chi^z+c_a^y c_a^z \chi^y\chi^z +
  (\frac{c_t \chi^0}{2c_0^2}-\frac{\chi^0}{\nu c_0})(\vec{c}_a\cdot\vec{\chi}) \\ 
   &+\frac{4}{15} 
  \big[(c_a^{x})^2(\chi^{x})^2+(c_a^{y})^2(\chi^{y})^2 +(c_a^{z})^2(\chi^{z})^2 \\ &-\frac{4}{\nu^2}(\vec{\chi}\cdot\vec{\chi})\big]  \bigg\} + \frac{c_0^2}{c_t^2} \epsilon_0 w_i \bigg\{ \frac{3}{2}(c^2 B^2+E^2)\\
  &+\frac{4}{5}[c^2 (\vec{B}\cdot \vec{B})+\vec{E}\cdot\vec{E}]-\frac{\nu}{\sqrt{3}}\big[(\vec{B}\times\vec{E})\cdot\vec{c}_a\big] \\
  &+\frac{\nu^2}{5}\big[ c^2(\vec{B}\cdot\vec{c}_a)^2 + (\vec{E}\cdot\vec{c}_a)^2 \big] -\frac{7 \nu^2}{20} \big[c_a^x c_a^y E^x E^y \\
  &+c_a^x c_a^z E^x E^z+c_a^y c_a^y E^z E^z)+ c^2(c_a^x c_a^y B^x B^y \\
  &+c_a^x c_a^z B^x B^z +c_a^y c_a^y B^z B^z)\big] \bigg\},
\end{aligned}
\end{equation}
where the second curly bracket is the contribution of the electromagnetic fields. Note that the second and third term in the first curly bracket are the contribution of the ideal gas equation of state which goes to zero in the ultrarelativistic limit ($\Gamma=4/3$ and $\epsilon \gg nc^2$) where the equation of state becomes $\epsilon=3p$.

As mentioned, Eq.\eqref{eq:rlb} can be used to solve the equations for the conservation of the total energy-momentum tensor, Eq.\eqref{Energy_momentum_cons}. To solve the equation for the conservation of number of particles, Eq.\eqref{Density_cons}, a separate distribution function based on the model proposed in Ref.\cite{hupp2011relativistic} is considered. Therefore, we add an extra
distribution function, $g_i$, which follows the dynamics of the Boltzmann
equation given by Eq.~\eqref{eq:rlb}, without the $\lambda_i$
coefficient term and by substituting $(\xi_\mu \chi^\mu/\nu^2)$ by unity. 
The corresponding modified equilibrium distribution function for the cell configuration given in Eq.\eqref{cellconfigs1} is given by:
\begin{equation}\label{Hupp}
g_i^{\rm eq}=w'_i n \gamma(u)\left(\frac{c_0}{c_t}+3(\vec{c_a}\cdot\vec{u})+\frac{9}{2}(\vec{c_a}\cdot\vec{u})^2-\frac{3}{2}u^2\right),
\end{equation}
with
\begin{equation}
w'_0=\frac{1}{10}, \quad w'_i=\frac{3}{10}-\frac{1}{6 c_a^2},
\end{equation}
for $1\leq i \leq6$, 
\begin{equation}
w'_i=\frac{1}{12 c_a^2}-\frac{3}{40},
\end{equation}
for $7 \leq i \leq 18$, where $w'_i$ are the respective discrete weights. Note that computing the macroscopic variables for the fluid, is not any more straightforward. We shall elaborate this issue later on.

\subsection*{Maxwell equations}

Now that the LB model for solving the fluid equations is discussed, let us explain our LB model for solving the governing equations of electromagnetic fields, i.e., Eqs. \eqref{Ampere}, \eqref{Faraday}, and \eqref{Current_cons}, with \eqref{Ohm} as Ohm's law. Our scheme is based on a 3D LB model for solving the Maxwell equations proposed in Ref.~\cite{mendoza2010three}, where several modifications are required to couple it to our solver of the fluid equations (mainly by modifying the distribution functions) as well as to use it for relativistic MHD (by using the relativistic Ohm's law). 

For this purpose, we use a cubic regular grid with 13 velocity vectors (D3Q13), where four auxiliary vectors are assigned to each of the vectors (two for the electric field and two for the magnetic field) for calculating the magnetic and electric fields. Note that, since Maxwell equations are first order, non self-interactive differential equations, the symmetry requirements are less strict than the symmetry requirement for the fluid dynamics equations. Therefore, compared to the lattices for fluid dynamics, less velocity vectors, e.g., 13 velocity vectors, are required. Indeed, one could also use the D3Q19 for the Maxwell equations, but it would unnecessarily spend more computational resources.

A simple streaming-collision evolution for the distribution function is considered as
\begin{equation}\label{LB_Maxwell}
h_{ij}^p(\vec{x}+\vec{v}_i^p \delta t,t+\delta t)-h_{ij}^p(\vec{x},t)=-\frac{1}{\tau_h}[h_{ij}^p(\vec{x},t)-h_{ij}^{p (\rm eq)}(\vec{x},t)],
\end{equation}
and
\begin{equation}
h_0^p(\vec{x},t+\delta t)-h_0^p(\vec{x},t)=-\frac{1}{\tau_h}[h_0^p(\vec{x},t)-h_0^{p (\rm eq)}(\vec{x},t)],
\end{equation}
where $h_{ij}^{p (\rm eq)}(\vec{x},t)$ and $h_0^{p (\rm eq)}(\vec{x},t)$ are the equilibrium distributions to be defined later. Here $i=0,1,2,3$ indicates the direction of the vectors, $p=0,1,2$ shows the plane where the vectors lie, and $j=0,1$ shows each of the two auxiliary vectors for the electric or magnetic field. Thus, there are four directions on three planes which gives 12 vectors, and including the rest vector, in total we have 13 vectors. These vectors (except the rest vector) lie on the diagonals of each plane and point to the edge-centers of a cube, so we can write the components as
\begin{equation}
\vec{v}_i^0=2c \left\{\cos{(2i+1)\pi/4}, \sin{(2i+1)\pi/4},0\right\},
\end{equation}
\begin{equation}
\vec{v}_i^1=2c \left\{\cos{(2i+1)\pi/4}, 0,\sin{(2i+1)\pi/4}\right\},
\end{equation}
\begin{equation}
\vec{v}_i^2=2c \left\{0,\cos{(2i+1)\pi/4}, \sin{(2i+1)\pi/4}\right\},
\end{equation}
in addition to the rest vector, i.e, $\vec{v}_0=(0,0,0)$.

\begin{figure}
\centering 
\includegraphics[width=0.7\columnwidth]{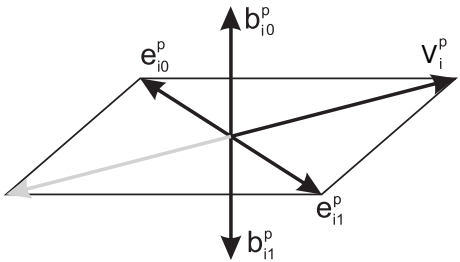} 
\caption{The configuration of the auxiliary vectors for the LB model to solve the Maxwell equations.}
\label{conexion}
\end{figure}

The distribution functions propagate with these vectors from cell to cell. Note that, unlike the LB models for fluid dynamics, these vectors do not represent the velocity of any particle. Associated to each velocity vector $\vec{v}_i^p$ there are two electric auxiliary vectors $\vec{e}_{ij}^p$ and two magnetic auxiliary vectors $\vec{b}_{ij}^p$, which are used to compute the electromagnetic fields. These vectors are perpendicular to $\vec{v}_i^p$. However, $\vec{e}_{ij}^p$ lies on the same plane as $\vec{v}_i^p$, while $\vec{b}_{ij}^p$ lies perpendicular to this plane. More accurately, we define them as (see Fig.\ref{conexion})
\begin{equation}
\vec{e}_{i0}^p=\frac{1}{2}\vec{v}_{[(i+3) \text{mod} 4]}^p, \quad \vec{e}_{i1}^p=\frac{1}{2}\vec{v}_{[(i+1) \text{mod} 4]}^p, 
\end{equation}
and
\begin{equation}
\vec{b}_{ij}^p=\frac{1}{2 c^2}\vec{v}_{i}^p \times \vec{e}_{ij}^p,
\end{equation}
where $(i)\text{mod} 4$ is a function that gives the remainder on the division of $i$ by $4$. To these vectors we shall add the null vectors, i.e., $\vec{e}_{0}=(0,0,0)$ and $\vec{b}_{0}=(0,0,0)$. This means that there are $13$ different electric vectors and $7$ different magnetic vectors.

In order to solve the Maxwell equations by the LB model we can write Ampere's law (Faraday's law) as time derivative of electric (magnetic) field plus the divergence of an antisymmetric tensor \cite{mendoza2010three}. We also consider the term$(-\mu_0 \vec{J})$ in Ampere's law (right hand side of Eq.\eqref{Ampere}) as an external force. Therefore, the macroscopic fields can be computed as
\begin{equation}\label{Cal_E}
\vec{E}'=\sum_{i=0}^{3}\sum_{p=0}^{2}\sum_{j=0}^{1}h_{ij}^p \vec{e}_{ij}^p,
\end{equation}
\begin{equation}\label{Cal_B}
\vec{B}=\sum_{i=0}^{3}\sum_{p=0}^{2}\sum_{j=0}^{1}h_{ij}^p \vec{b}_{ij}^p,
\end{equation}
and 
\begin{equation}\label{Cal_rhoc}
\rho_c=h_0+\sum_{i=0}^{3}\sum_{p=0}^{2}\sum_{j=0}^{1}h_{ij}^p.
\end{equation}
Note that the effect of the external force still needs to be considered to get the correct electric field and $\vec{E}'$ is the electric field before considering the external force.

It can be shown that to recover the Maxwell equations, for the current model, $\tau_h=\frac{1}{2}$ should be considered. Unlike the LB models for fluid dynamics, this value for the relaxation time does not lead to numerical instabilities, because of the linear nature of the Maxwell equations. For the case of $\tau_h=\frac{1}{2}$ the external force in Ampere's law can be included in a rather simple way, and $\vec{E}$ becomes (see Ref.~\cite{mendoza2010three})
\begin{equation}\label{Ex_force}
\vec{E}=\vec{E}'-\frac{\delta t}{2} \mu_0 c^2 \vec{J},
\end{equation}
where, according to the Ohm's law, Eq.\eqref{Ohm}, $\vec{J}$ is a function of $\vec{E}$. By substituting Ohm's law in Eq.\eqref{Ex_force} we obtain a system of three equations and three unknowns ($E^x$, $E^y$ and $E^z$), which can be solved analytically. Having the values of each component of the electric field, $\vec{J}$ can be calculated using Ohm's law, consequently. More discussion about computing the macroscopic variables shall be provided later.

The discretized equilibrium distribution function to recover the correct Maxwell equations reads as follows 
\begin{equation} \label{Maxwell-eq}
h_{ij}^{p(\rm eq)}(\vec{x},t)=\frac{1}{16} \vec{v}_i^p \cdot \vec{J}+\frac{1}{8 c^2}\vec{E}\cdot\vec{e}_{ij}^p+\frac{1}{8}\vec{B}\cdot\vec{b}_{ij}^p,
\end{equation}
and 
\begin{equation}
h_0^{\rm eq}(\vec{x},t)=\rho_c.
\end{equation}
A Chapman-Enskog expansion shows that the current model recovers Ampere's law, Faraday's law and current conservation, Eq.\eqref{Current_cons}. The latter follows from the evolution of $h_0$.

The divergence free condition for the magnetic field, Eq.\eqref{Gauss_B}, can be treated as a constraint on the initial condition, since by taking the divergence of Faraday's law one can show that the time derivative of $\vec{\nabla} \cdot \vec{B}$ is always zero. Therefore, if $\vec{\nabla} \cdot \vec{B}=0$ is set for the initial condition it will hold for later times as well. The same is true for the Gauss law, Eq.\eqref{Gauss_E}, using Eqs. \eqref{Ampere} and \eqref{Current_cons}. 

\subsection*{Coupling between fluid and electromagnetic fields}

Having explained the appropriate solvers for fluid equations and Maxwell equations, we next discuss how to compute the macroscopic variables. As mentioned, the model of Anderson-Witting is used for the collision term in the solver for the equation of conservation of energy-momentum. Hence, according to the Landau-Lifshitz decomposition \cite{cercignani} the macroscopic variables can be calculated by solving an eigenvalue problem resulting from multiplying the relation for the energy-momentum tensor by the covariant four-vector velocity. Using the definition of the total energy-momentum tensor, Eqs.\eqref{El_total},\eqref{EM_Fluid} and \eqref{EM_EM} with the help of the relation $F^{\alpha \beta} F_{\alpha \beta}=2(c^2 B^2-E^2)$ we get
\begin{equation}
U_\mu \left[T^{\mu \nu}-\epsilon_0 F^{\mu\rho}F^\nu_\rho\right]=\left[\epsilon+\frac{\epsilon_o}{2}(c^2 B^2-E^2)\right] U^\nu,
\end{equation}
Here, $\left[\epsilon+\frac{\epsilon_o}{2}(c^2 B^2-E^2)\right]$ and $U^\nu$ are the largest eigenvalue and corresponding eigenvector of the tensor $\left[T^\mu_\nu-\epsilon_0 F^{\mu \rho} F_{\nu \rho}\right]$, respectively. The total energy -momentum tensor can be calculated using the relation $T^{\mu \nu}=\sum_{i=1}^{19} p_i^\mu p_i^\nu f_i$. On the other hand, the tensor $F^{\mu \rho} F_{\nu \rho}$ depends on $\vec{E}$ and $\vec{B}$. As mentioned, for calculating $\vec{E}$ the value of the external force, which depends on the velocity, is required. However, the value of the velocity is not yet computed. To solve this problem, we use the value of the electromagnetic fields from the previous time step to calculate the tensor $\left[T^\mu_\nu-\epsilon_0 F^{\mu \rho} F_{\nu \rho}\right]$. The largest eigenvalue and corresponding eigenvector (velocity) can be calculated numerically using the power method. Knowing the velocity, one can calculate the density using the first order moment relation, i.e., $N^{\mu}=n U^\mu=\sum_{i=1}^{19} p_i^\mu g_i$. After that, $\vec{B}$ is calculated using Eq.\eqref{Cal_B}, where, the $h_{ij}^p$ are obtained using the LB equation for the Maxwell equation, i.e, Eq.~\eqref{LB_Maxwell}, in which the $h_{ij}^{p \rm (eq)}$ (Eq.~\eqref{Maxwell-eq}) are known from the previous time step. Having the velocity and magnetic field, $\vec{E}$ can be computed using Eq.\eqref{Cal_E} and including the external force as described before. Thus, it is easy to compute $\vec{J}$ using Ohm's law. Considering the eigenvalue $\left[\epsilon+\frac{\epsilon_o}{2}(c^2 B^2-E^2)\right]$, it is possible to compute $\epsilon$ and through the equation of state Eq.\eqref{EOS} one can compute $p$. Finally, $\rho_c$ can be calculated directly from Eq.\eqref{Cal_rhoc}. All the $13$ unknown variables for each cell can be computed in this way. Note that using the values of electromagnetic fields from the previous time step to calculate the tensor, leads to an error which goes to zero as the time step ($\delta t$) decreases. In fact, in the next section our numerical results show that the error is ignorable.

Finally, since two separate solvers for fluid equations and Maxwell equations are used in this model, we need to make sure that the time evolution of the two solvers are consistent with each other, i.e. the distribution functions for the fluid and the electromagnetic solvers must evolve simultaneously in time. For the electromagnetic solver the time evolution, i.e. $\delta t$ depends on the value of the spatial spacing, i.e., $\delta x$ through the relation $\delta x/\delta t=\sqrt{2} c$, whereas for the fluid solver $\delta x/\delta t$ can be chosen freely, as long as the stability of the numerical model is not impaired. Since the value of $\delta x$ is the same for both solvers, to ensure the same value of $\delta t$ for both solvers, the numerical value of the velocity of light is adjusted accordingly, and used for both solvers. Also, note that, in order to describe signals that move at the speed of light, e.g. electromagnetic waves, the numerical quasi-particles in the LB model should move faster than light (according to the relation $\delta x/\delta t=\sqrt{2} c$). Nevertheless, in the continuum limit the differential equations that our model reproduces, which characterize the physical system, never violate the principle of relativity.

\section {Test simulations and applications}
\label{Test simulations and application} 

In this section, we present some numerical tests in order to validate our numerical model along with some applications for the resistive relativistic MHD. More specifically, test simulations for the propagation of Alfv\'en waves and the evolution of a self-similar current sheet are considered, and as applications for the resistive relativistic MHD LB model, we study the magnetic reconnection driven by the Kelvin-Helmholtz instability and present the results of a 3D simulation of magnetic reconnection in a stellar flare due to the shear velocity in the photosphere. The purpose of the first test simulation (Alfv\'en wave), is to validate the numerical method in the limit of ideal MHD, while the second test is to validate the model recovering the correct dynamics in the resistive regime. In the following simulations numerical units are used and $\mu_0=1$ is considered. 

\subsection*{Propagation of Alfv\'en wave}
\label{Propagation of Alfven wave}

\begin{figure}
\centering 
\subfigure []{\includegraphics[width=1\columnwidth]{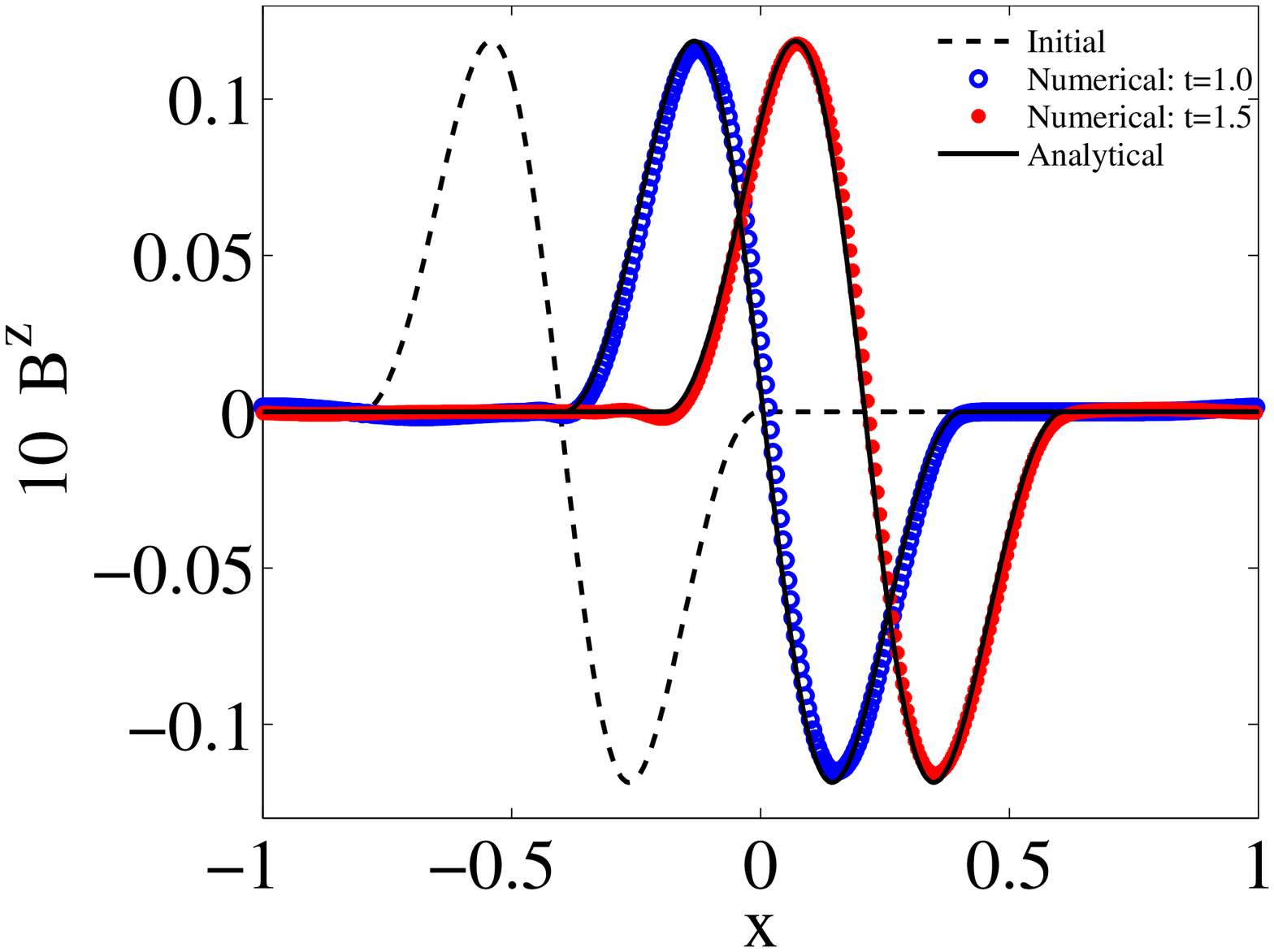} \label{Alfven_bz}}
\subfigure []  {\includegraphics[width=1\columnwidth]{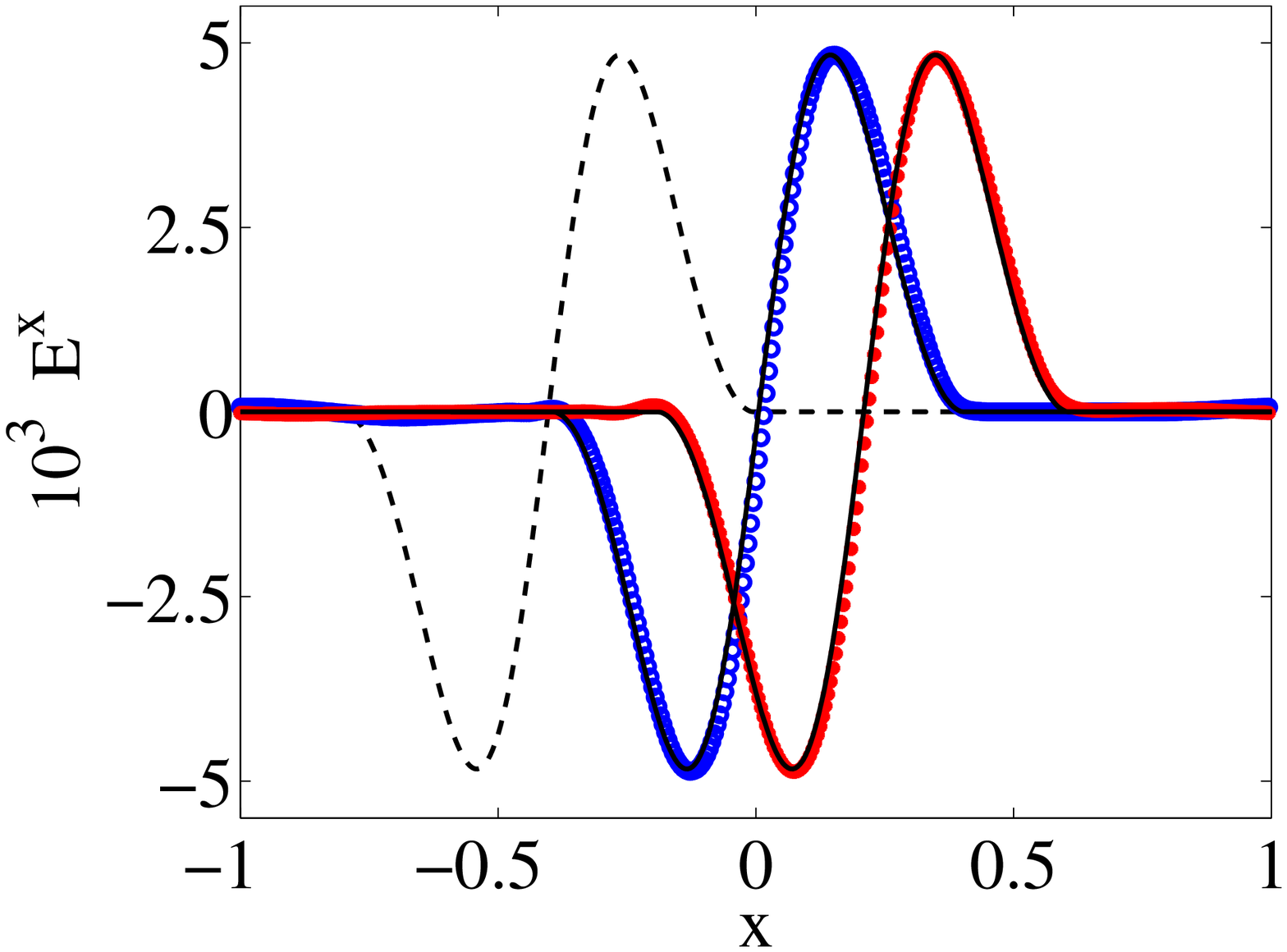}\label{Alfven_ex}}
\caption{(Color Online) Results of the numerical test for the propagation of Alfv\'en waves in the limit of ideal MHD ($\sigma=10^5$) for (a) magnetic field in $z$-direction and (b) electric field in $x$-direction. For both plots, the numerical results are shown at $t=1$ (blue circle symbols) and at $t=1.5$ (red square symbols). The dashed black line shows the initial condition and solid black lines are the exact analytical solutions at the corresponding time.}
\label{Alfven}
\end{figure}

This test deals with the propagation of an Alfv\'en wave along a uniform background field in the limit of ideal MHD. The initial condition is the same as in Ref.\cite{komissarov2007multidimensional}. We set $n=1.0$, $p=1.0$, $B^x=B_0=1.0$ and $B^y=0.1$ and we consider a one dimensional domain defined in the range of $-1\leq x \leq 1$, where the initial wave is located at $x_0<x<x_1$, with $x_0=-0.8$ and $x_1=0$. Outside the region of the initial wave it is assumed that $\vec{u}=\vec{0}$ and $B^z=0$. Inside the region of initial wave we have  
\begin{equation}
B^z=\eta_A B_0 \sin\left[2\pi (3 x_*^2 -2x_*^3) \right],\quad x_*=\frac{x-x_0}{x_1-x_0},
\end{equation}
and 
\begin{equation}
u^z=-\frac{v_A}{B_0} B^z, \quad u^x=u^y=0,
\end{equation}
 where the Alfv\'en velocity is defined as
\begin{equation}
\begin{aligned}
v_A^2 &=\frac{2 B_0^2 c^2}{\epsilon+p+B_0^2(1+\eta_A^2)} \\
& \times \left[ 1+\sqrt{1-\left( \frac{2\eta_A B_0^2}{\epsilon+p+B_0^2(1+\eta_A^2)} \right)^2} \right]^{-1}.
\end{aligned}
\end{equation}
Here we use the value of $\eta_A=0.118591$ which gives $v_A=0.40785$. Ohm's law for ideal MHD, i.e. Eq.\eqref{Ohm_ideal} is used to calculate the initial electric field and Eq.\eqref{Gauss_E} to measure the initial value of $\rho_c$. To achieve the ideal regime a very high conductivity ($\sigma=10^5$) is considered. The equation of state, Eq.\eqref{EOS}, with $\Gamma=4/3$ is used. The domain is discretized using $400$ cells and we set $\delta x/\delta t=\sqrt{2}$, which gives $c=1$. The value of $\tau$ for the fluid LB model is set to $1$ and $\alpha=0.1$. Open boundary conditions for each cell at the left and right boundaries are implemented by copying the distribution functions from the neighboring cell in the direction perpendicular to the boundary.

The simulation runs until $t=1.5$ and the results are presented in Fig.\ref{Alfven}. In the limit of ideal MHD the generated wave should travel with the Alfv\'en velocity without any distortion. Here we have compared our numerical results with the exact analytical solution of the ideal MHD at two different times, i.e., $t=1.0$ and $t=1.5$. The results are presented for the magnetic field in $z$-direction (Fig.\ref{Alfven_bz}) and the electric field in $x$-direction (Fig.\ref{Alfven_ex}). Fig.\ref{Alfven} shows that at each of the considered times and for both $B^z$ and $E^x$, very good agreement is observed between the numerical and analytical results. This simulation takes $\sim 320$ ms on a single core of an Intel CPU with $2.40$ GHz clock speed. Note that, the analytical results are obtained by simply shifting the initial wave by the Alfv\'en velocity. Apart from validating the numerical method, this test shows the ability of the model to deal with high conductivity (low resistivity) regimes, recovering the ideal MHD limit.

\subsection*{Evolution of self-similar current sheets}
\label{Evolution of self-similar current sheet}

After validating the model for the ideal MHD case, we consider here a test problem in the resistive case for which the evolution of a current sheet is investigated. We assume that the magnetic pressure ($B^2/2$) is much smaller than the plasma pressure ($p$), so that the fluid is not affected by the evolution of the current sheet and changes in the magnetic field. We know that when the magnetic field changes its sign within a thin layer a current sheet forms. Thus, for our case, we assume that the magnetic field has only a tangential component $\vec{B}=(0,B^y,0)$, where $B^y=B(x,t)$ changes sign within a thin current sheet of width $\Delta l$. If the fluid is set initially to equilibrium, by considering a constant pressure in the domain, the evolution of the current sheet becomes a diffusion process. By assuming that the diffusion time-scale is much longer than the light propagating time-scale, we can neglect the displacement current ($\partial_t \vec{E}$) in Ampere's law. In the rest frame, by inserting the relation $\vec{J}=\sigma \vec{E}$ in Ampere's law, using Faraday's law, and plugging in the mentioned one-dimensional magnetic field of the current sheet, one gets
\begin{equation}
\partial_t B^y-\frac{1}{\sigma} \partial_x^2 B^y=0.
\end{equation}
As the diffusion process continues and the width of the current sheet becomes much larger than the initial width ($\Delta l$), the expansion becomes self-similar and the analytical solution has the form
\begin{equation}\label{Current_sheet}
B(x,t)=B_0 \text{erf} \left( \frac{1}{2}\sqrt{\frac{\sigma}{t}} x \right),
\end{equation}
where $B_0$ is the magnetic field outside of the current sheet and $\text{erf}$ is the error function. The above equation describes the evolution of a current sheet, providing $\partial_t \vec{E}$ is ignorable.

\begin{figure}
  \centering 
 \includegraphics[width=1\columnwidth]{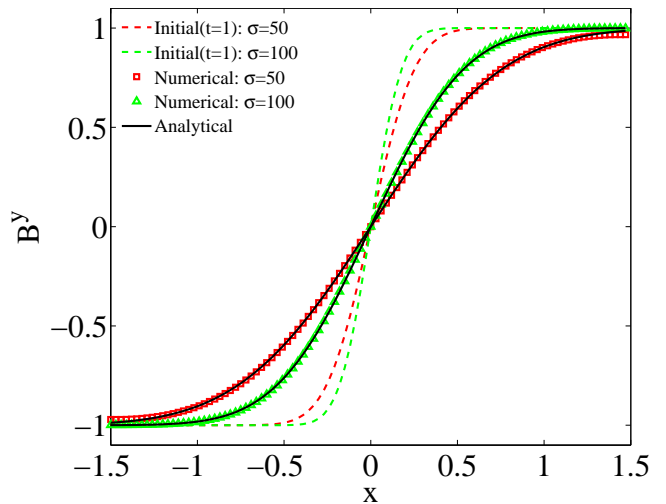}
\caption{(Color Online) Results of the numerical test for the evolution of a current sheet in the resistive regime. Dashed green line (upper curve at $x>0$) and dashed red line (second upper curve at $x>0$) show the initial condition of the current sheet for $\sigma=100$ and $\sigma=50$, respectively, which corresponds to the analytical solution, Eq.\eqref{Current_sheet}, at $t=1$. The green triangle and red square symbols show the result of the numerical solution at $t=9$, for $\sigma=100$ and $\sigma=50$, respectively. The solid black lines show the exact analytical solution at this time.}
\label{Cur_sheet}
\end{figure}

For the numerical test a domain of $-1.5\leq x\leq 1.5$ is discretized using $100$ cells, where open boundary conditions are considered for the left and right boundaries. The initial values $p=50$, $n=1$, $\vec{u}=\vec{E}=\vec{0}$ and $B^x=B^z=0$, are considered, and the initial $B^y$ is computed using Eq.\eqref{Current_sheet} at $t=1$ with $B_0=1$. In the numerical model $\Gamma=4/3$, $\delta_x/\delta t=\sqrt{2}$, $\tau=1$ and $\alpha=0.1$ are considered. The simulation runs until $t=8$ and the results are compared to the analytical results at $t=9$ (since the initial condition is assumed to be at $t=1$). Two values of uniform conductivity, $\sigma=100$ and $\sigma=50$, are considered and the results of the comparison with the analytical solution for both cases along with the initial conditions are presented in Fig.\ref{Cur_sheet}. One can see that the numerical and analytical results are almost indistinguishable and that the current sheet in a domain with $\sigma=50$ diffuses faster than the domain with $\sigma=100$ due to the higher resistivity. This test validates the resistive part of the numerical model and shows the capability of the model for simulating resistive problems far from the ideal MHD limit. The above simulation, for $\sigma=100$, takes $\sim 70$ ms on a single core of an Intel CPU with $2.40$ GHz clock speed.

\begin{figure}
  \centering 
 \includegraphics[width=0.95\columnwidth]{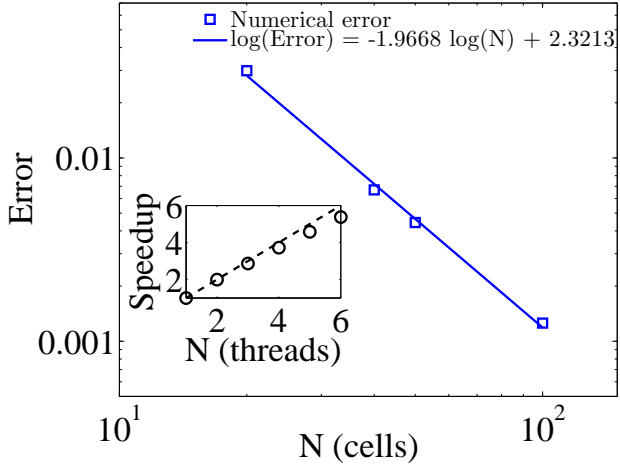}
\caption{Results of the convergence test for the simulation of the evolution of a current sheet in the resistive regime with $\sigma=100$ in a log-log plot. Symbols show the numerical error and the solid blue line is the fitted line with slope $-1.9668$, which shows a second order convergence. In the inset the speedup obtained by using the openMP parallelizing method is reported versus the number of threads. The dashed line shows the ideal speedup. }
\label{error}
\end{figure}

To check the convergence of the model, we implement the same current sheet simulation with $\sigma=100$ for different grid resolutions. Fig.\ref{error} reports the error versus the number of cells in a log-log plot, and the slope of the fitted line (blue solid line) shows that the model is second order as we expect for a lattice Boltzmann scheme. The error is calculated as follows:
\begin{equation}
E_N=\left(\frac{1}{N} \sum_{i=1}^N (B^y_{\rm num}-B^y_{\rm num(200) })^2  \right)^{1/2},
\end{equation}
where $N$ is the number of cells, and $B^y_{\rm num}$ and $B^y_{\rm num(200)}$ are the numerical results for the cases with $N$ and $200$ grid resolution, respectively. Due to the fact that the analytical expression, i.e., Eq.~\eqref{Current_sheet}, is only approximately accurate, a fine mesh ($200$ cells) is used as a reference to calculate the error. In other words, for each grid resolution, the numerical result at each position is compared to the numerical results of the fine resolution. Additionally, as explained in the introduction, one of the advantages of using lattice Boltzmann methods is its simplicity and efficiency on parallel computers. To show that this also holds for our model, we use a simple openMP parallelizing method and simulate the 3D extension of the aforementioned current sheet problem with $150\times150\times150$ cells and $\sigma=100$, until $5$ time steps. The preliminary resulting speedup for a few number of threads is shown in the inset of Fig.\ref{error}, where one can see that even for a straight forward parallelizing method, a satisfactory level of efficiency is achieved. However, for a thorough study of the parallelizing efficiency, a much larger number of threads and more advanced parallelizing methods should be experimented, which is an interesting topic for future research.

\subsection*{Magnetic reconnection driven by Kelvin-Helmholtz instability}
\label{Magnetic reconnection driven by Kelvin-Helmholtz instability}

\begin{figure}
\centering 
\subfigure [] {\includegraphics[width=0.48\columnwidth]{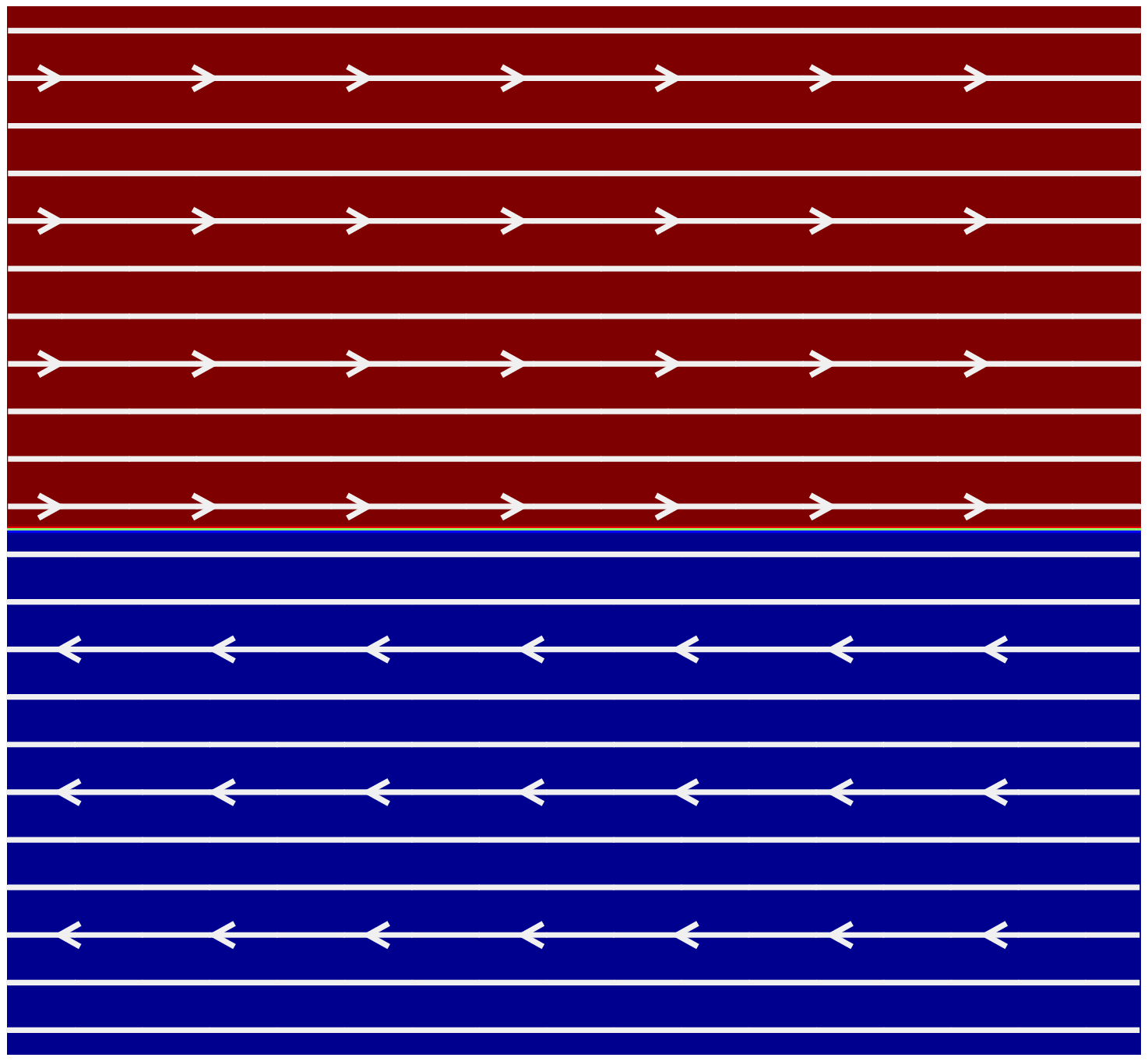}\label{90_t_0}}
\subfigure [] {\includegraphics[width=0.48\columnwidth]{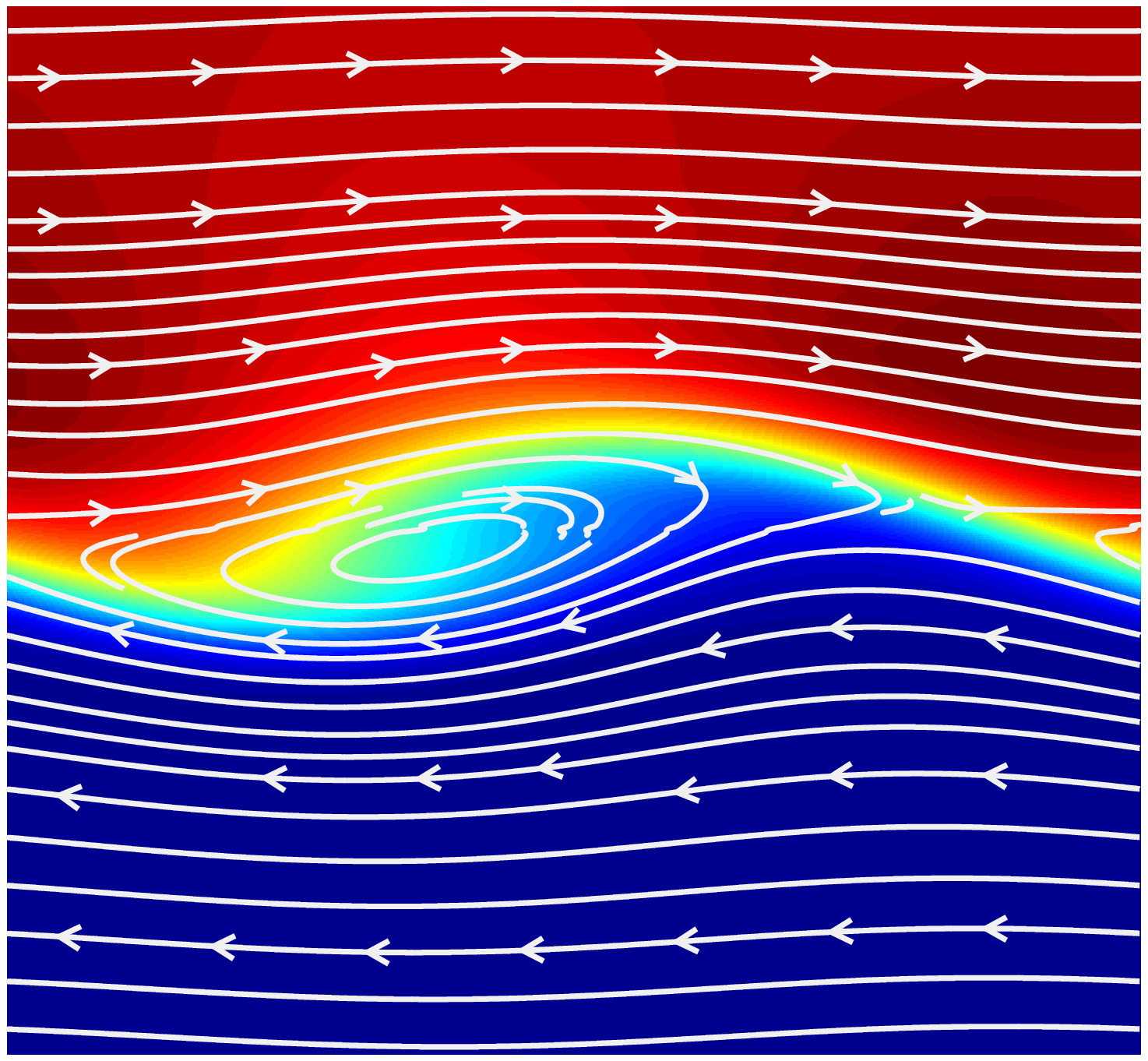}\label{90_t_3}}
\subfigure [] {\includegraphics[width=0.48\columnwidth]{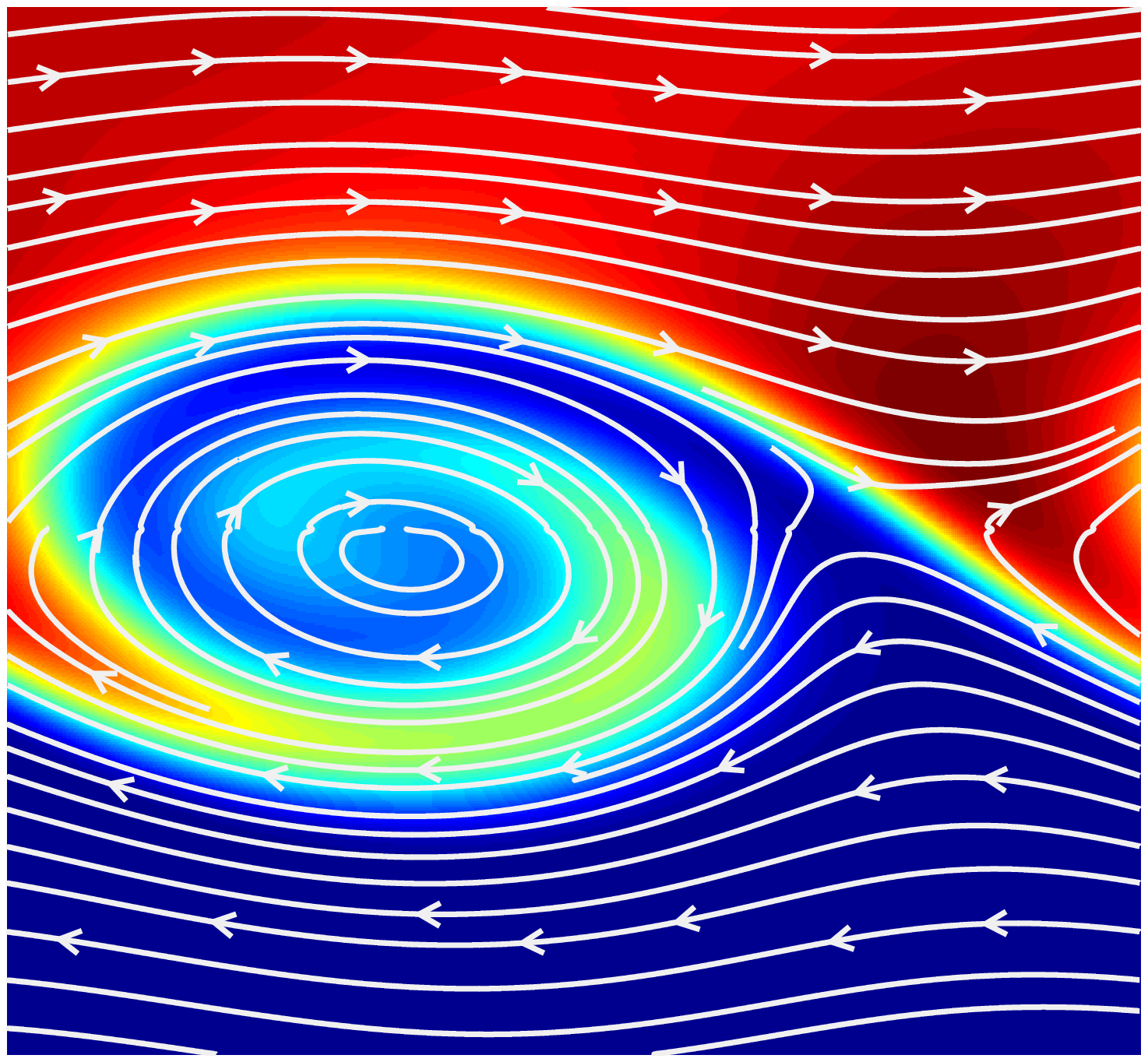}\label{90_t_6}}
\subfigure [] {\includegraphics[width=0.48\columnwidth]{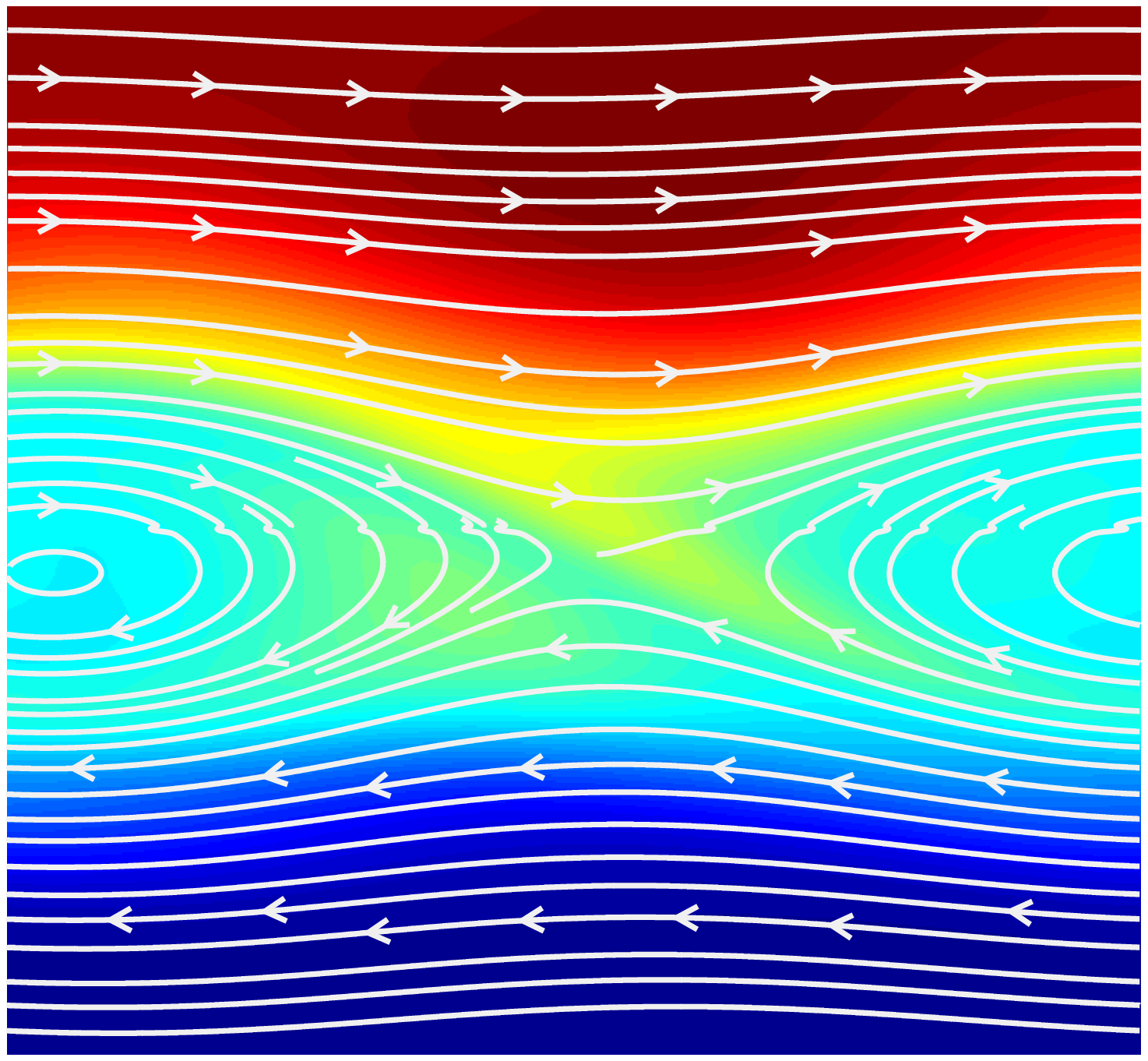}\label{90_t_15}}
\subfigure{\includegraphics[width=0.1\columnwidth]{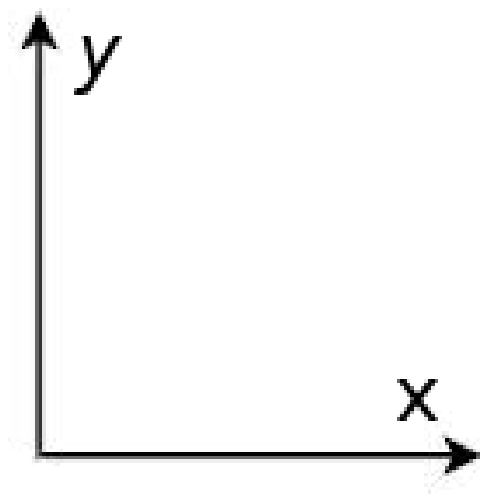}}
\subfigure{\includegraphics[width=0.9\columnwidth]{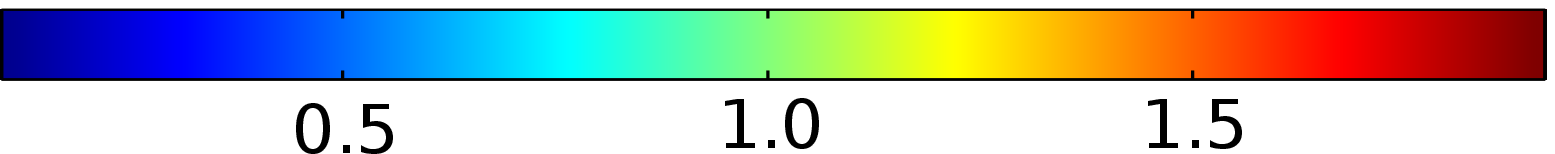}}
\caption{(Color Online) Snapshots of the density for the KH instability at times (a) $t=0.0$ (initial condition) (b) $t=3.31$ (c) $t=6.62$ (d) $t=15.47$, for the case with $\sigma=100$, $\Delta n=1.8$, $U_0=0.6 c$. The white lines show the magnetic field lines.}
\label{KH_90}
\end{figure}

After validating the numerical model, we now study the magnetic reconnection process driven by the Kelvin-Helmholtz (KH) instability. As mentioned before, the magnetic reconnection is a process where magnetic lines change their topology. At the place where the magnetic lines reconnect, usually a null point forms where the magnetic field vanishes. There are several theoretical descriptions for the magnetic reconnection including the well known Sweet-Parker \cite{sweet1958neutral,parker1957sweet} and Petschek \cite{petschek1963magnetic} models.

\begin{figure}
\centering 
\includegraphics[width=1\columnwidth]{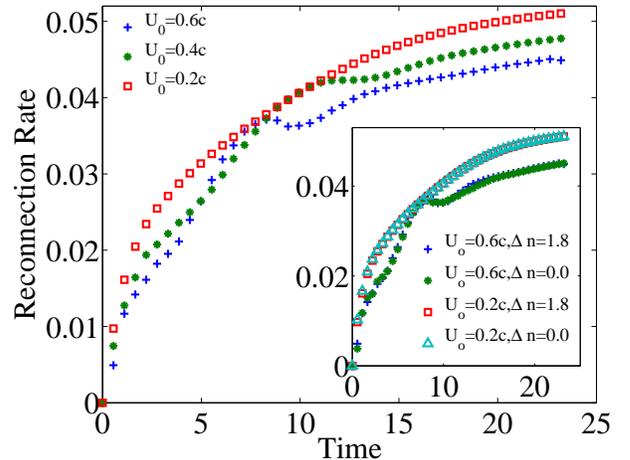}
\caption{(Color Online) Reconnection rate versus time for the case $\sigma=100$, $\Delta n=1.8$ for different values of $U_0$. In the inset the results are shown for two different values of $\Delta n=1.8$ and $\Delta n=0$ for $\sigma=100$.}
\label{R_t_U0_n}
\end{figure}

The Sweet-Parker model is based on the discussion of pressure balance in the reconnection region, where the reconnection region is assumed to be dominated by diffusion while the outside region is assumed to be ideal. Due to the magnetic diffusion, the plasma is driven into the current sheet (inflow) with velocity $v_{\rm in}$ in the direction perpendicular to the length of the current sheet. The conversion of magnetic energy within the current sheet thrusts the plasma out with the velocity $v_{\rm out}$ in the direction of the length of the current sheet. It is shown that the reconnection rate $R$, which is defined as the ratio between $v_{in}$ and $v_{\rm out}$ is proportional to $S^{-\frac{1}{2}}$, where $S=\mu_o L v_A \sigma$ is the Lundquist number (magnetic Reynolds number), with $L$ as characteristic length. The reconnection rate in this model is usually small due to the high aspect ratio of the reconnection region (which is proportional to the inflow and outflow velocities assuming the incompressibility condition).

In the Petschek model, it is assumed that the magnetic energy can be liberated not only in the current sheet but also as pairs of slow shocks which stem from the edge of the sheet. Therefore, the reconnection region can be smaller than the one for the Sweet-Parker model, which can lead to higher reconnection rate. In this model $R$ is proportional to $(\ln S)^{-1}$.

As discussed, magnetic reconnection is believed to have prominent effects in many high energy astrophysical events and therefore it should be strongly influenced by relativistic effects. Although the mechanism of relativistic reconnection is not well understood, recent theoretical and numerical studies of the relativistic Sweet-Parker and Petschek models show the same proportionality relation between $R$ and $S$ \cite{lyubarsky2005relativistic,takahashi2011scaling}. In the numerical simulations, it is important how one triggers the reconnection process. If a local increase in the conductivity is used to trigger the reconnection, Petschek type reconnection is observed with ``x-type'' null point, while when a perturbation in the magnetic potential is applied, Sweet-Parker type reconnection is observed with ``y-type'' null point \cite{zenitani2010resistive}. This is similar to the non-relativistic numerical results.
Here instead of using either of the mentioned ways, we use a hydrodynamic instability to trigger the reconnection. In particular, the KH instability is chosen because of its wide range of applications in astrophysical events as discussed in the introduction section. Therefore, two dimensional simulations of the KH instability in a Harris like current sheet are performed for different values of shear velocity, density ratio and conductivity. A domain of $-0.5\leq x\leq 0.5$ by $-0.5\leq y\leq 0.5$ is discretized using $512 \times 512$ cells. The following initial conditions are considered:
\begin{equation}
u_x=\frac{U_0}{2}\tanh \left(\frac{y}{a} \right), \quad n=n_0+\frac{\Delta n}{2} \tanh \left(\frac{y}{a} \right),
\end{equation}
where $U_0$ defines the shear velocity, $\Delta n=(n_{\rm up}-n_{\rm down})$ is the density difference between the upper ($y>0$) and lower  ($y<0$) part of the domain, $n_0=1$, $p=1$ and $a=\delta x$ is considered. Furthermore, 
\begin{equation}
B^x=B_0 \tanh \left(\frac{y}{a} \right),
\end{equation}
while $B^y=B^z=0$ and $B_0=0.06$ is considered. Note that high values of magnetic field will stabilize the KH instability due to the magnetic tension of
the background magnetic field. The KH instability is triggered by a perturbation in the velocity in $y$-direction, namely
\begin{equation}
u_y=u_{\rm pert} \sin \left(k x \right) \exp \left(- \frac{y^2}{b^2} \right),
\end{equation}
with $u_{\rm pert}=0.01$ as the perturbation amplitude, $k=2 \pi$ for a single mode perturbation, and $b=10 \delta x$. For the left and right boundaries ($x=\pm0.5$) periodic boundary conditions are considered while for the upper and lower boundaries ($y=\pm0.5$) open boundary conditions are implemented. The simulation is performed for different values of $U_0=0.6c,0.4c,0.2c$ and $\sigma=100,80,60,40,20$ for two values of $\Delta n=1.8$ and $\Delta n=0$ where the latter corresponds to the case with initial uniform density. For the numerical simulation $\Gamma=4/3$, $\delta x/\delta t=2.5 \sqrt{2}$, $\tau=1$ and $\alpha=0.1$ are considered. The initial electric field can be simply computed using Faraday's law (dropping the time derivative term because of the stationary state) and Ohm's law, knowing the fact that the initial velocity and magnetic fields are in the same plane. Additionally, for the open boundary condition, and to ensure the divergence free condition of the magnetic field, the normal component of the magnetic field is adjusted in order to have $\vec{\nabla}\cdot \vec{B}=0$ at the boundaries. This will slightly improve the numerical stability of the model. Note that for the electric field, the same idea (setting the normal electric field on the boundaries according to $\vec{\nabla}\cdot \vec{E}=\rho_c$) does not affect the numerical stability. Therefore, for each cell, the open boundary condition for the electric field is implemented like for the other variables, i.e., by copying the distribution functions from the neighboring cell in the direction perpendicular to the boundary.
\begin{figure}
\centering 
\subfigure [] {\includegraphics[width=0.48\columnwidth]{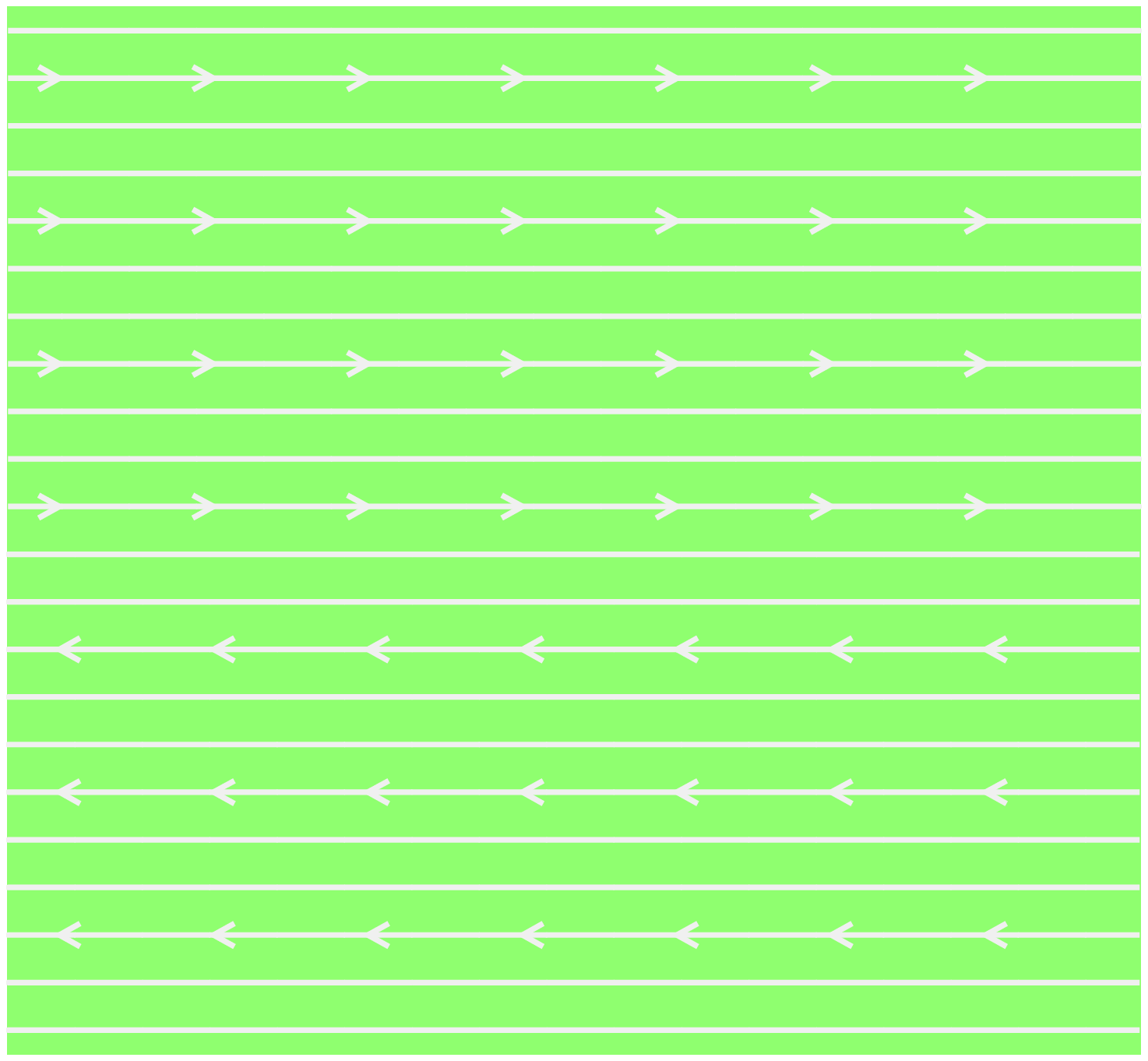}\label{0_t_0}}
\subfigure [] {\includegraphics[width=0.48\columnwidth]{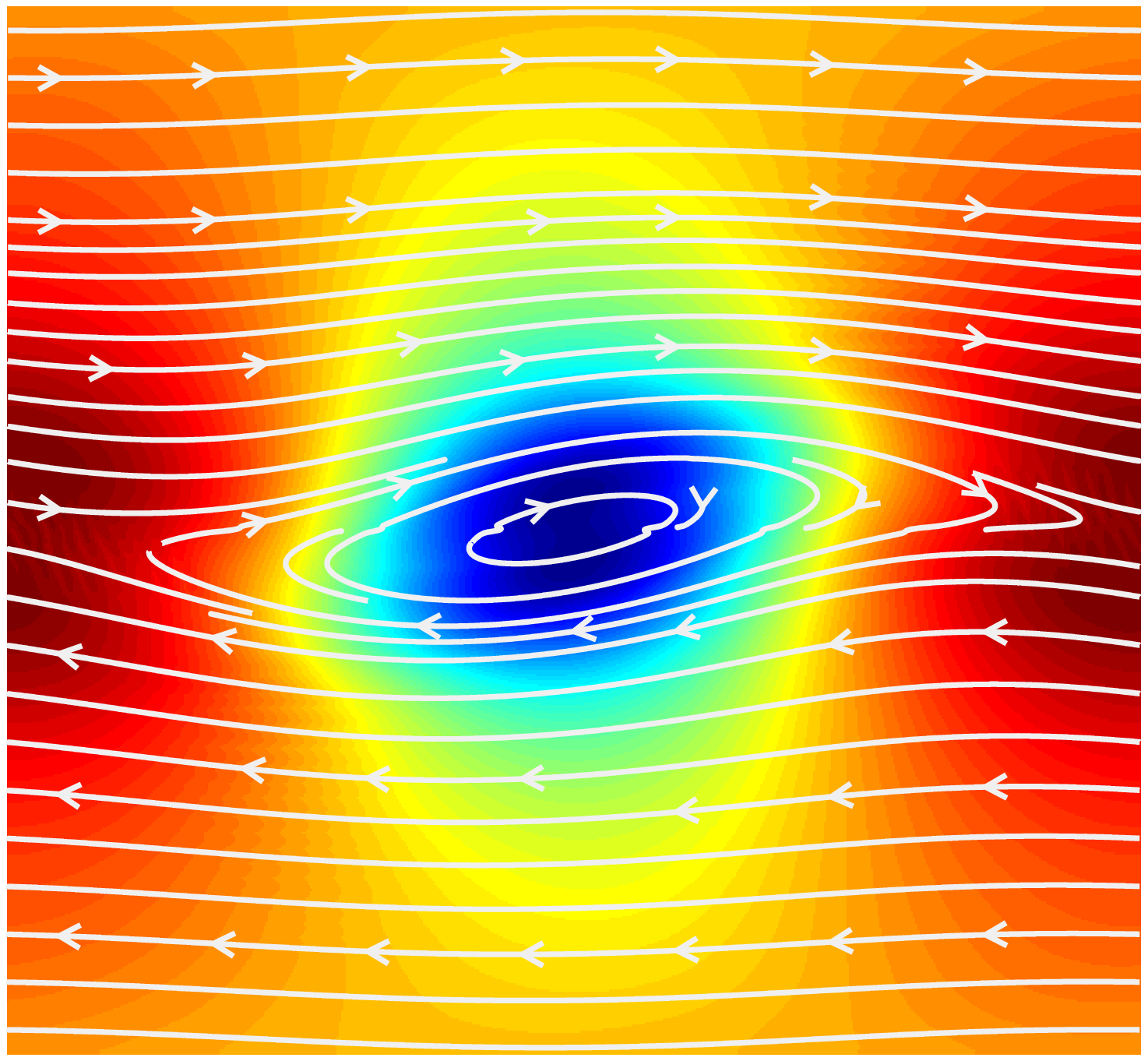}\label{0_t_3}}
\subfigure [] {\includegraphics[width=0.48\columnwidth]{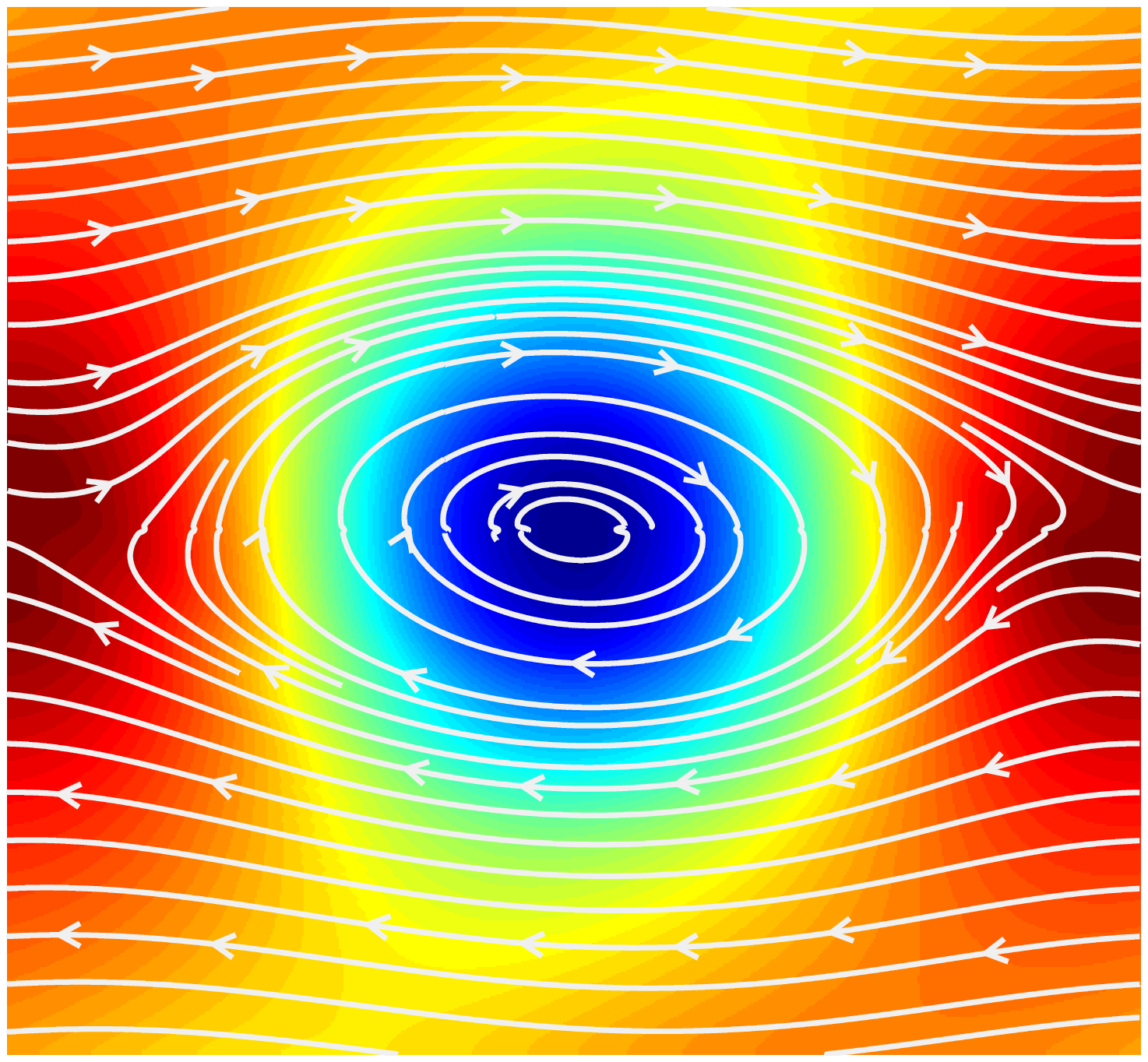}\label{0_t_6}}
\subfigure [] {\includegraphics[width=0.48\columnwidth]{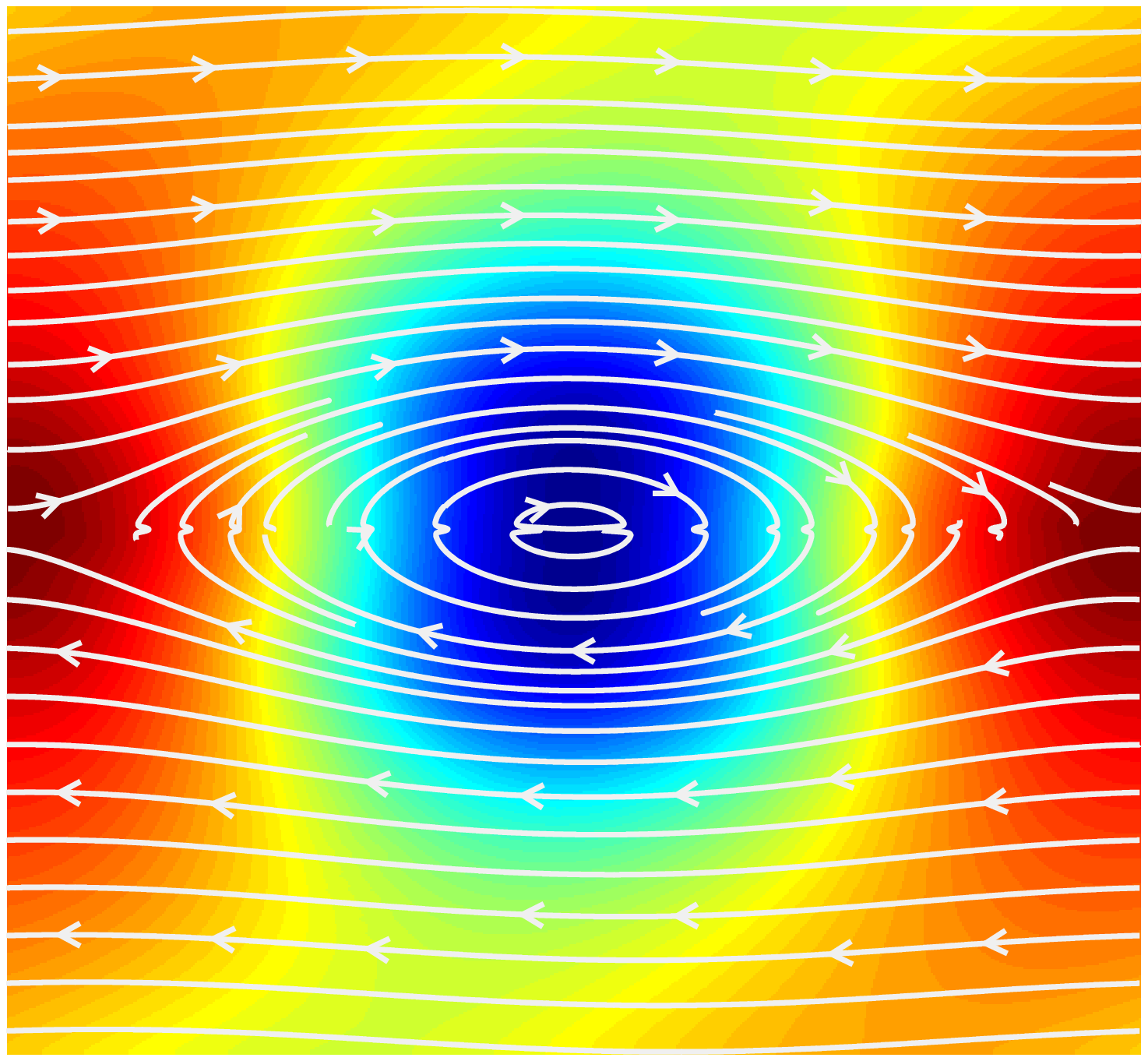}\label{0_t_15}}
\subfigure{\includegraphics[width=0.1\columnwidth]{xy_axis.eps}}
\subfigure{\includegraphics[width=0.9\columnwidth]{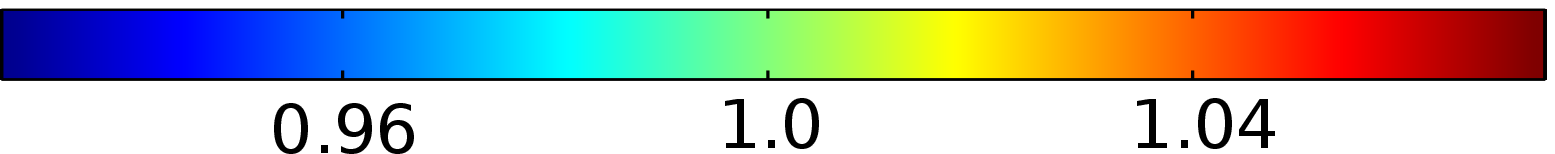}}
\caption{(Color Online) Snapshots of the density for the KH instability at times (a) $t=0.0$ (initial condition) (b) $t=3.31$ (c) $t=6.62$ (d) $t=15.47$, for the case with $\sigma=100$, $\Delta n=0$, $U_0=0.6 c$. The white lines show the magnetic field lines.}
\label{KH_0}
\end{figure}

The snapshots of the density for the case with $\sigma=100$, $\Delta n=1.8$ and $U_0=0.6c$ are presented in Fig.\ref{KH_90} for different time steps. One can see the evolution of the instability in time where after an initial linear growth, a non-linear stage takes place which leads to the penetration and mixing of the lighter and heavier fluids, where the characteristic structure of the KH instability forms. The reconnection is triggered by the instability and occurs in the initial current sheet, where the magnetic field changes sign. As the instability evolves, the location of the reconnection null point changes. Since no external force is implemented, the instability finally smoothens out due to the dissipation in the system.

\begin{figure}
\centering 
\includegraphics[width=1.0\columnwidth]{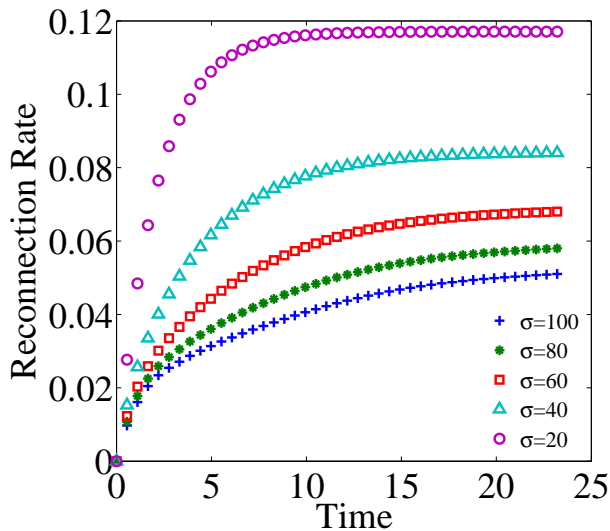}
\caption{(Color Online) Reconnection rate versus time for the case $\Delta n=1.8$ and $U_0=0.2c$ for different values of $\sigma$.}
\label{R_t_sigma}
\end{figure}

We are interested in the reconnection rates of the considered cases. Since the initial condition is not in the rest frame, it is not practical to find the reconnection rate based on the definition mentioned before (ratio between inflow and outflow velocities). Instead we use \cite{zenitani2009two}
\begin{equation}
R(t)=\left(\frac{\Delta E^z}{B_0}\right)_{\rm null},
\end{equation}
i.e., the generated out-of-plane component of the electric field ($\Delta E^z=E^z-E^z_0$, $E^z_0$ being the initial out-of-plane electric field) normalized by $B_0$ at the null point, which shows the rate at which the magnetic flux is convected to this point (see Ref.\cite{johnson2013gaussian}). Here we investigate the effects of different parameters on the reconnection rate. First we study the effects of the parameters related to the KH instability, namely $U_0$ and $\Delta n$. Fig. \ref{R_t_U0_n} shows the results for the reconnection rate versus time for different values of $U_0$, when $\sigma=100$ and $\Delta n=1.8$. It is shown that an increase in the magnitude of the shear velocity reduces the reconnection rate. This is due to the stabilizing effects of the shearing velocity on the tearing instability \cite{chen1997tearing}, an MHD instability that appears in connection with sheared magnetic field. Also note that the bumps that appear in the reconnection rate at high shear velocities is due to the transition of the instability into its stationary state, where one can appreciate that, for higher shear velocities, this transition occurs at earlier times.

Additionally, in the inset of Fig.\ref{R_t_U0_n}, for the case with $\sigma=100$, the effects of different values of $\Delta n$ is shown. One can notice that changing the value of $\Delta n$ from $1.8$ to $0$ (initially uniform density) has negligible effects on the reconnection rate, for different magnitudes of the shear velocity. This is despite the fact that the hydrodynamics of the system for the initially uniform density is quite different from the case with inhomogeneous densities. Fig.\ref{KH_0} shows the snapshots of the density for the case $\Delta n=0$, $\sigma=100$ and $U_0=0.6c$, where the well known ``cat's eye'' structure of KH instability for the case with initially uniform density can be recognised. Comparison between Fig.\ref{KH_90} and Fig.\ref{KH_0} shows that, for the case with the initially uniform density, the results are symmetric and because of the form of the initial perturbation, the location of the null point is always at the boundary, unlike the previous case, where the location of the null point changes with time. 

\begin{figure}
\centering 
\includegraphics[width=1\columnwidth]{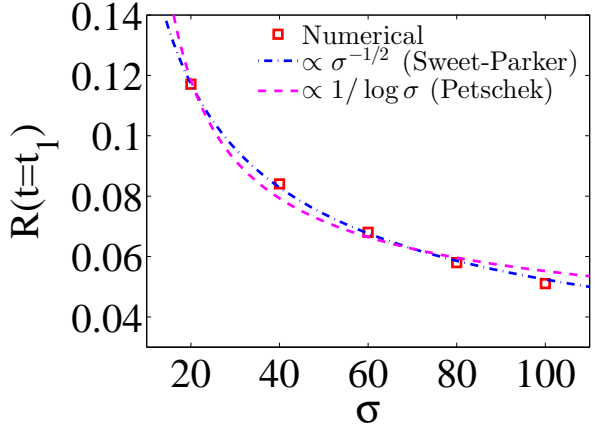}
\caption{(Color Online) $R(t)$ at time $t=23.2$ for different values of $\sigma$ based on the numerical results (red symbols). Best fitting curves using the proportionality relation for the Sweet-Parker (dashed-dotted blue) and Petschek (dashed red) models are shown.}
\label{R0_sigma}
\end{figure}

Another parameter that we are interested in studying is the conductivity $\sigma$. Fig.\ref{R_t_sigma} shows the results of the reconnection rate versus time for the case with $U_0=0.2c$ and $\Delta n=1.8$ for different values of the conductivity. As shown, the reconnection rate increases faster in time and reaches a higher value for lower conductivities (higher resistivity). The fact that the reconnection rate increases by increasing the resistivity is also expressed in the model of Sweet-Parker and Petschek. The interesting point is to inspect the exact relation between the reconnection rate and the resistivity. As mentioned before, in the Sweet-Parker model, $R$ is proportional to $\sigma ^{-\frac{1}{2}}$ (assuming constant Alfv\'en velocity) and in the Petschek model $R$ is proportional to $(\ln \sigma)^{-1}$. To compute this proportionality relation for our results, the reconnection rates $R(t=t_1)$ at the final time $t_1=23.2$ are used. The results are shown in Fig.\ref{R0_sigma}, where, as expected, $R_0$ decreases with increasing $\sigma$. The blue dashed line, and the red dashed-dotted line in Fig.\ref{R0_sigma} are the best fitting curves for the data using the proportionality relations suggested by Sweet-Parker and Petschek models, respectively. One can see that the results clearly do not follow the Petschek scaling law while match the Sweet-Parker scaling law very closely. 

\subsection*{Three-dimensional magnetic reconnection in a stellar flare}
\label{Three-dimensional magnetic reconnection in a stellar flare}
In this section, we show the results of the 3D numerical simulation of magnetic reconnection in a stellar flare, which is driven by a shear velocity on its photosphere. Thus, a domain of $-3 \leq x \leq 3$ by $-3 \leq y \leq 3$ by $0 \leq z \leq 6$ is discretized by $256\times256\times256$ cells. The configuration of the initial condition is chosen to mimic the arcade and the flux rope of a stellar flare \cite{chen2000emerging}. The total potential field (flux function) is defined as
\begin{equation}
\psi=\psi_b+\psi_l+\psi_i.
\end{equation}
This configuration consists of a background magnetic field that is produced by four line electrical currents (just below the photosphere), which determines $\psi_b$ and an image electrical current (below the photosphere), which determines $\psi_i$. This background magnetic field yields a null point above the photosphere. Additionally, a line electrical current contained within the flux rope with finite radius, which determines $\psi_l$, is added to the null point. Note that if we do not consider the flux rope, the magnetic field configuration is a potential quadrupole field.

\begin{figure}
\centering 
\subfigure [] {\includegraphics[trim=10mm 10mm 10mm 10mm, clip, width=0.49\columnwidth]{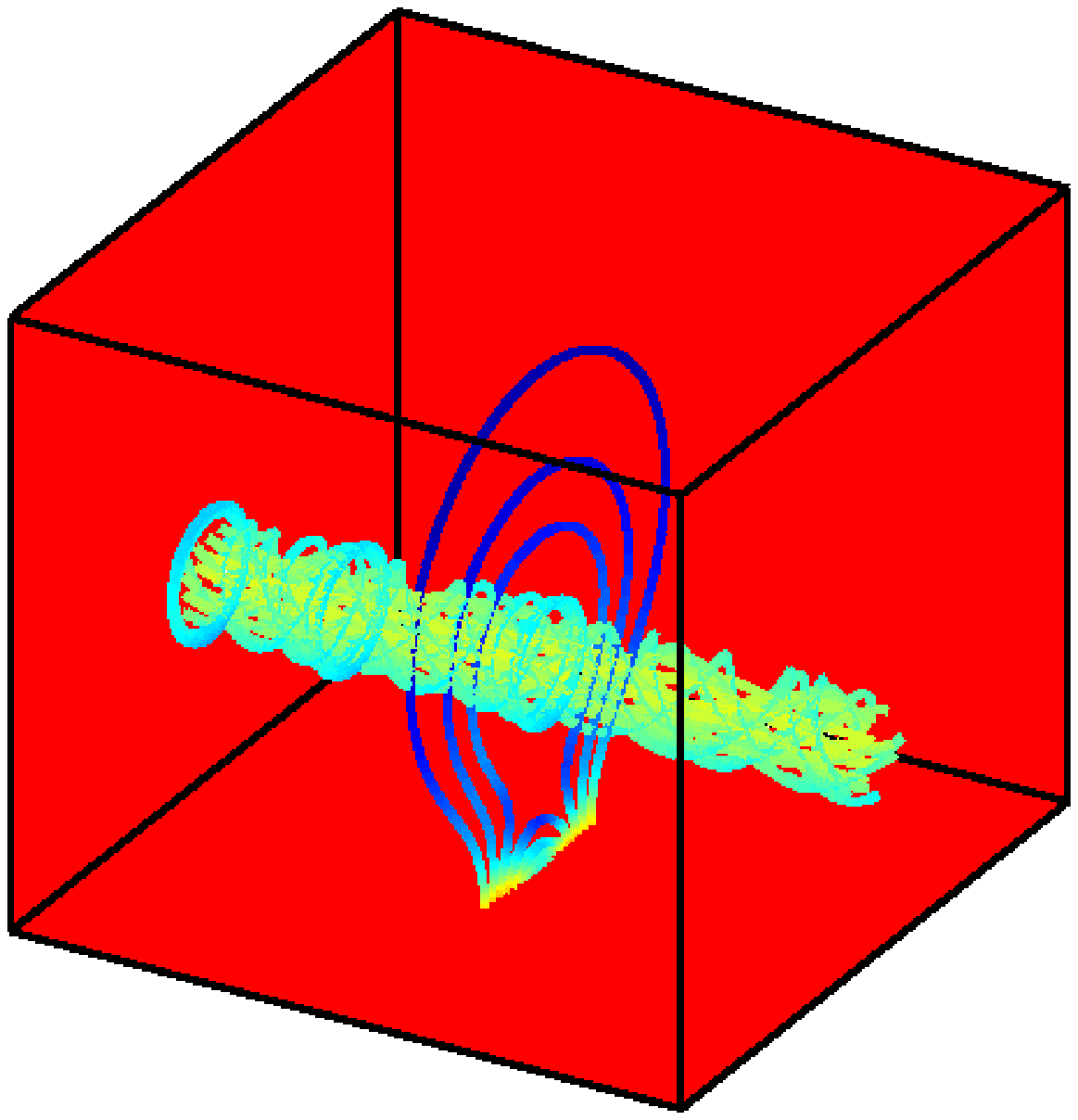}\label{solarqp_t_1}}
\subfigure [] {\includegraphics[trim=10mm 10mm 10mm 10mm, clip, width=0.49\columnwidth]{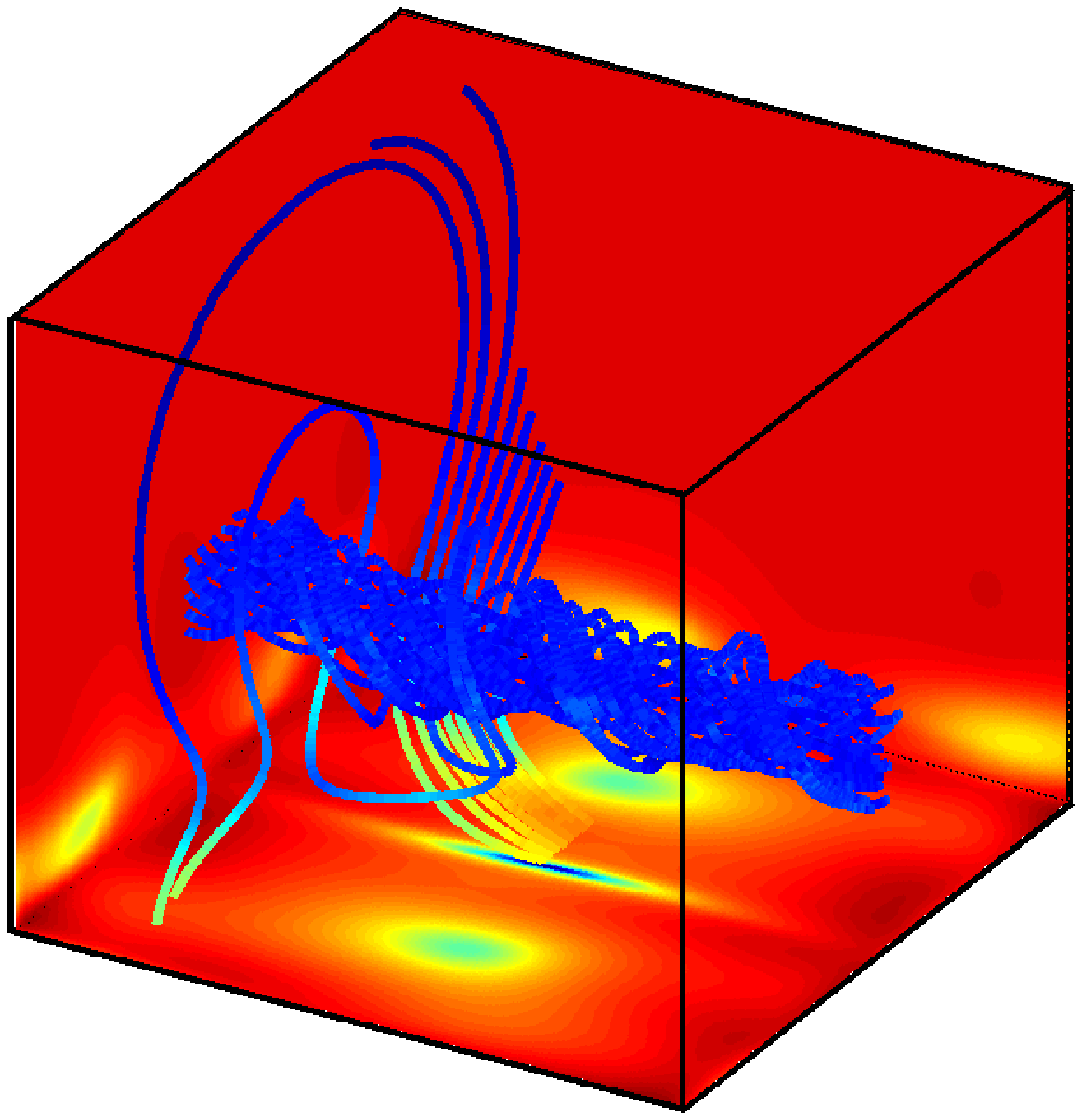}\label{solarqp_t_6}}
\subfigure [] {\includegraphics[trim=10mm 10mm 10mm 10mm, clip, width=0.49\columnwidth]{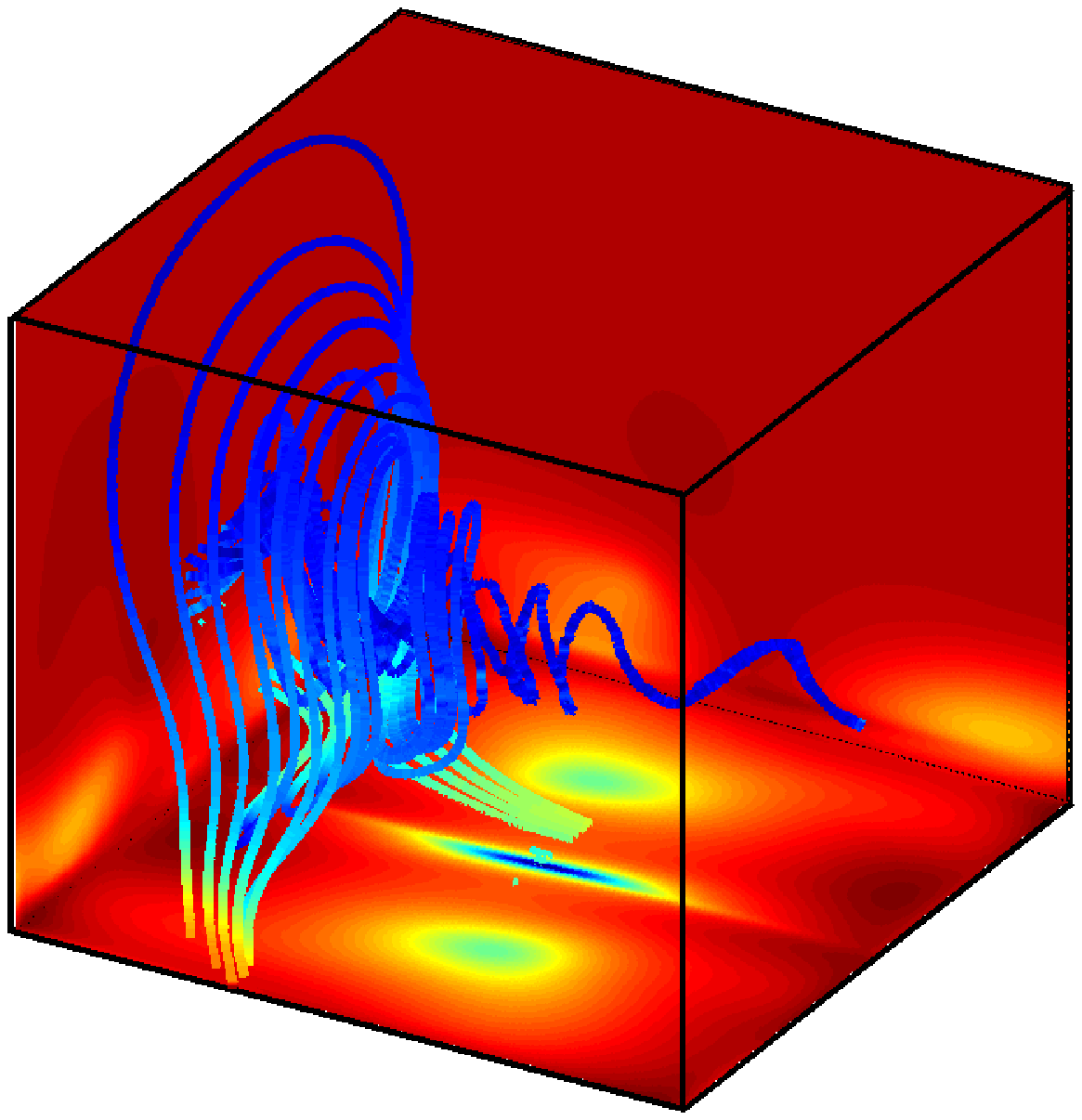}\label{solarqp_t_13}}
\subfigure{\includegraphics[width=0.15\columnwidth]{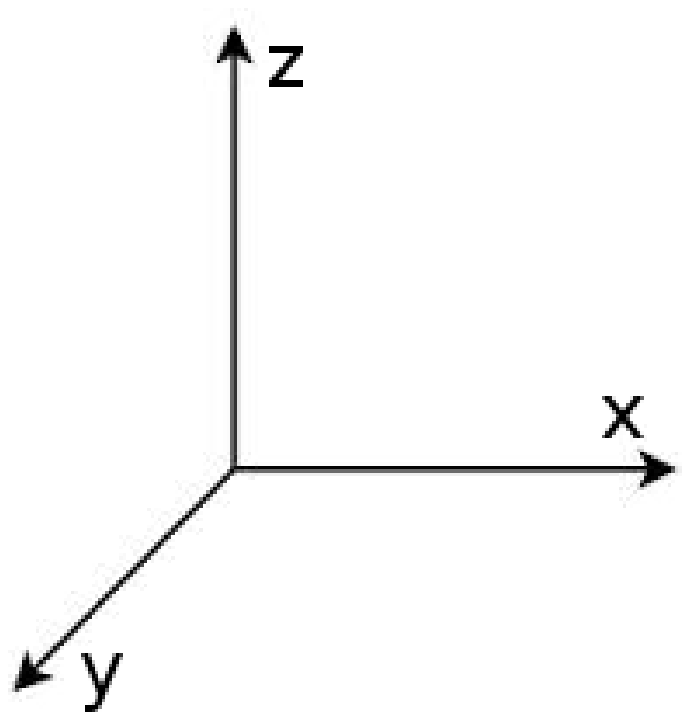}}
\subfigure{\includegraphics[width=0.9\columnwidth]{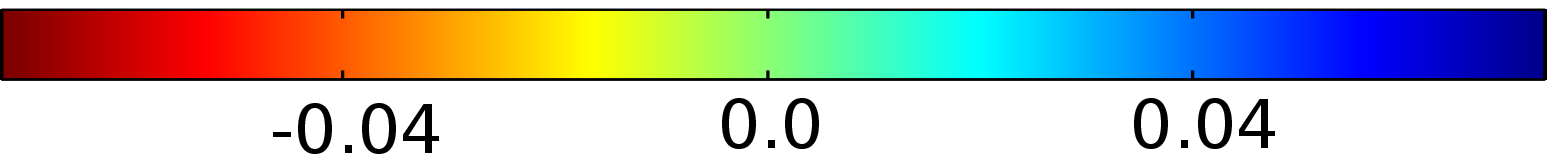}}
\caption{(Color Online) Snapshots of the 3D magnetic reconnection in a stellar flare due to the shear flow at times (a) $t=0$ (initial condition), (b) $t=19.86$  and (c) $t=43.02$. The colors of the magnetic lines show the magnitude of the magnetic field, where blue to red show low to high values. On the outer boundary surfaces of the domain the colors indicate the value of the vorticity which is indicated in the colorbar.}
\label{3DKH}
\end{figure}

The potential fields generated by these currents are defined as follows:
\begin{equation}
\begin{aligned}
&\psi_b=\\
&c_b\log{\frac{[(x+0.3)^2+(z+0.3)^2][(x-0.3)^2+(z+0.3)^2]}{[(x+1.5)^2+(z+0.3)^2][(x-1.5)^2+(z+0.3)^2]}},
\end{aligned}
\end{equation}   
\begin{equation}
\psi_i=-\frac{r_0}{2}\log{[x^2+(z+h)^2]},
\end{equation} 
\begin{equation}
\psi_l=\left\{ \begin{array}{ll}
\frac{r^2}{2r_0}, & r\leq r_0,\\
\\
\frac{r_0}{2}-r_0\log{r_0}+r_0\log{r}, & r> r_0,
\end{array} \right.
\end{equation}
where $r=[x^2+(z-h)^2]^{1/2}$ is the distance from the center of the flux rope, $h$ is the height of the flux rope, which is set to $2.0$, $r_0$ is the radius of the flux rope, which is set to $0.5$, and $(\pm 1.5, -0.3)$ and $(\pm 0.3, -0.3)$ are the $(x,z)$ positions of the four line currents. Here, $c_b$ represents the strength of the background magnetic field, which is set to $0.2534$ \cite{chen2000emerging}. The resulting magnetic field is calculated by taking the curl of the total potential field, i.e., $\vec{B^*}=\vec{\nabla}\times\psi \hat{e_y}$, and in order to adjust the magnitude of the magnetic field, each component of the magnetic field is multiplied by $(B_0/B^*_{\rm max})$, where $B^*_{\rm max}$ is the maximum of the computed magnetic field in the domain and $B_0=0.06$ is considered. To satisfy the force balance inside the flux rope, a magnetic component should be added in the direction perpendicular to the magnetic field, i.e., $y$-direction, which has the following form:
\begin{equation}
B^y=\left\{ \begin{array}{ll}
(B_0/B^*_{\rm max})\sqrt{2\left( 1-\frac{r^2}{r_0^2}\right)}, & r\leq r_0,\\
\\
0, & r> r_0.
\end{array} \right.
\end{equation} 
One can see the illustration of the initial configuration of the magnetic field in Fig.~\ref{solarqp_t_1}. The initial velocity is set to zero, and the initial density and pressure are considered to be uniform and equal to $1.0$ everywhere in the domain. In order to compute the initial values for the electric field and electrical current density, the numerical code is used in an iterative process where the value of magnetic field and velocity is fixed at each time-step. The converged values of the electric field and electrical current provide the initial condition for these variables. Open boundary conditions are considered for all the boundaries, and as mentioned before, the normal component of the magnetic field is adjusted on each boundary to satisfy the divergence free condition of the magnetic field.

\begin{figure}
\centering 
\subfigure [] {\includegraphics[trim=10mm 10mm 10mm 10mm, clip,width=0.49\columnwidth]{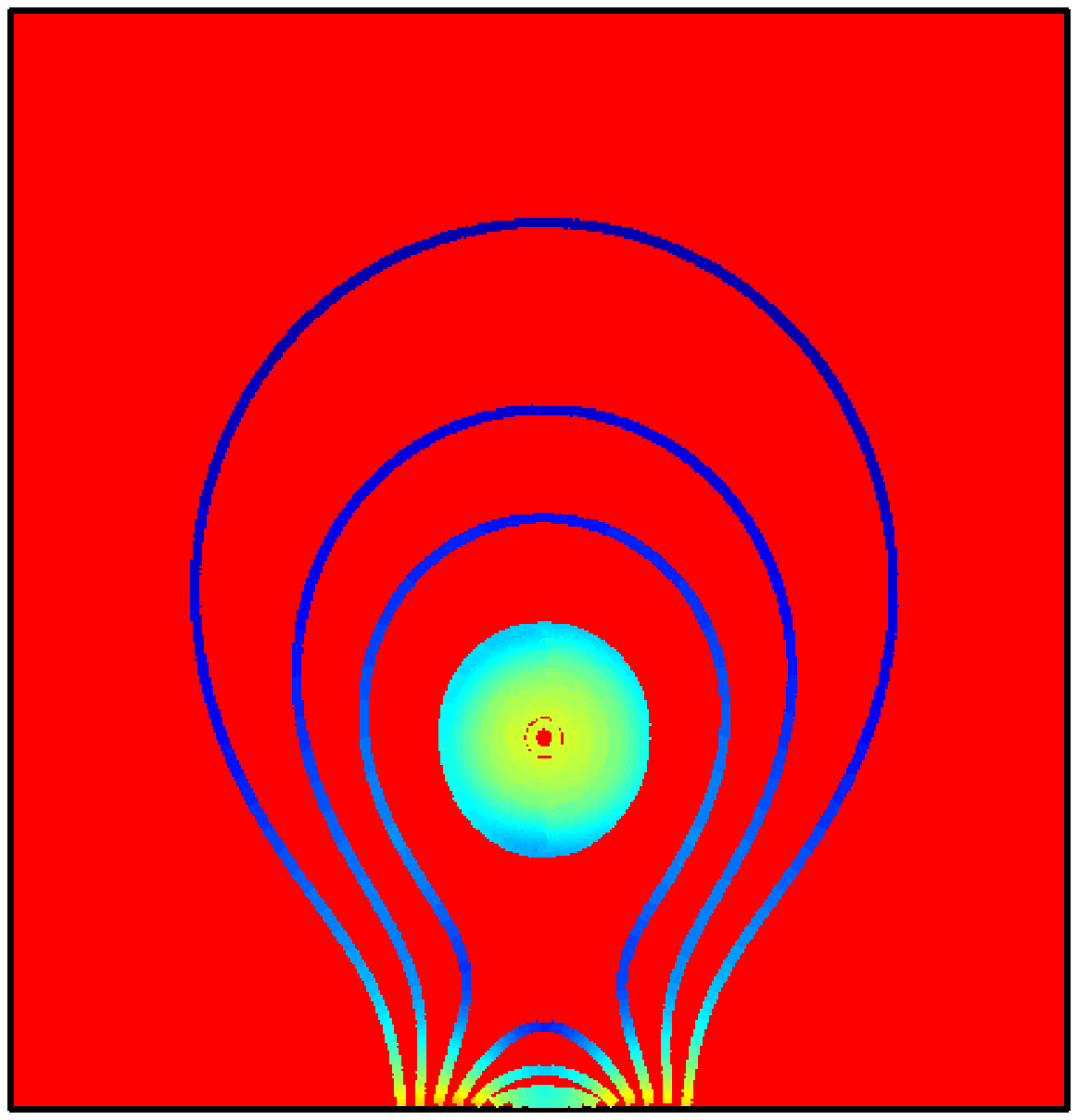}\label{solarqp_xz_t_1}}
\subfigure [] {\includegraphics[trim=10mm 10mm 10mm 10mm, clip,width=0.49\columnwidth]{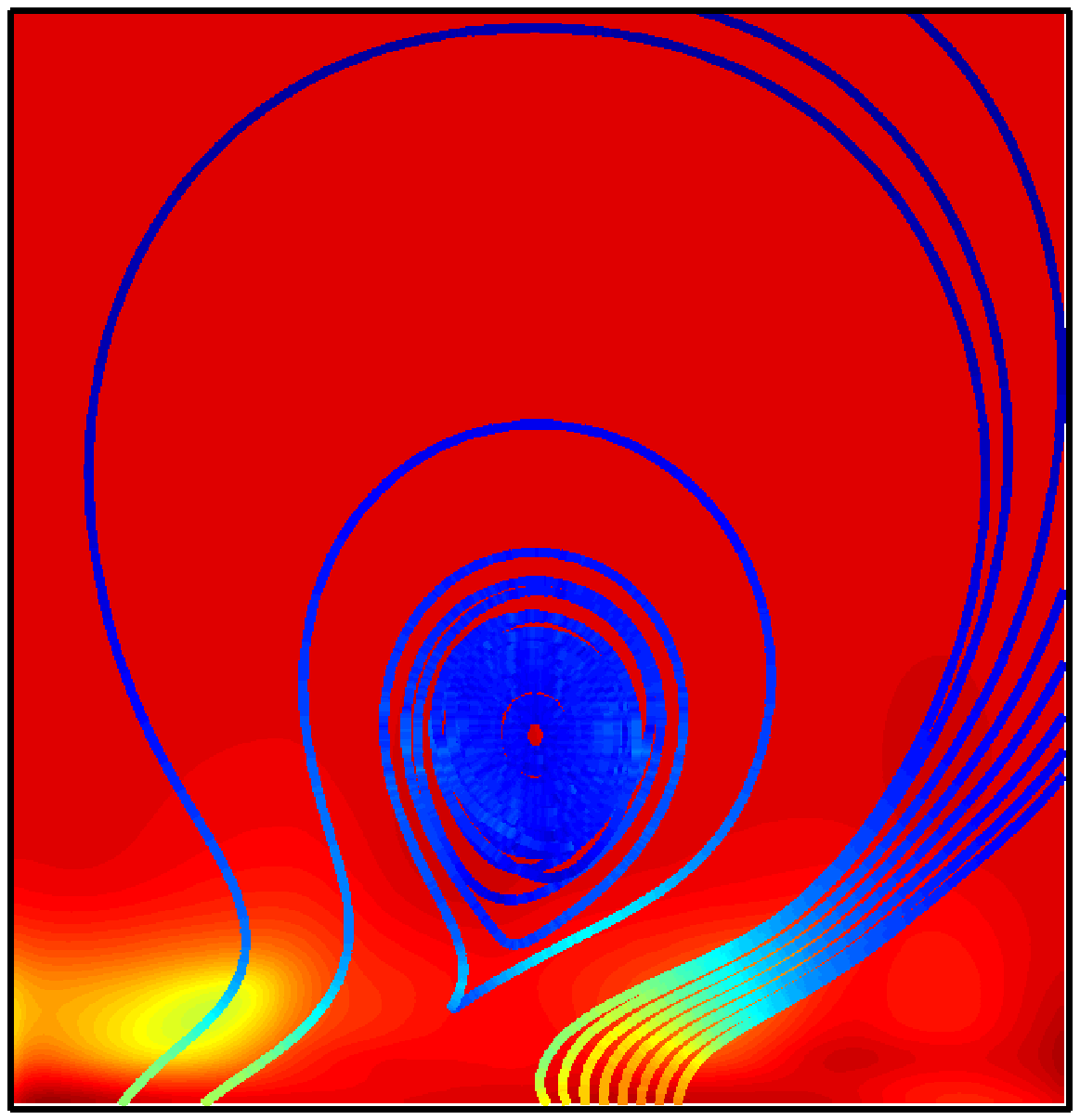}\label{solarqp_t_xz_6}}
\subfigure [] {\includegraphics[trim=10mm 10mm 10mm 10mm, clip,width=0.49\columnwidth]{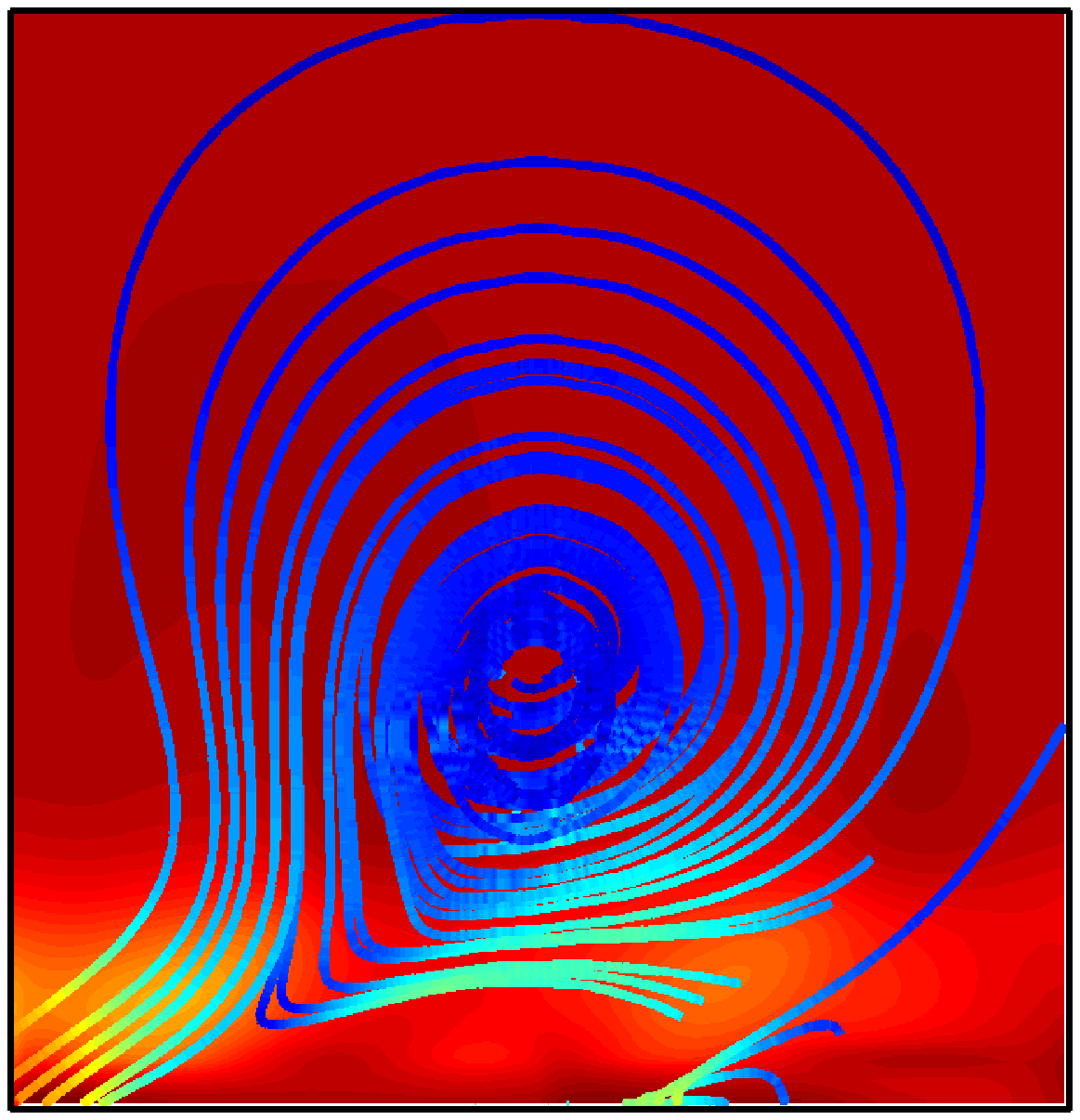}\label{solarqp_t_xz_13}}
\subfigure{\includegraphics[width=0.1\columnwidth]{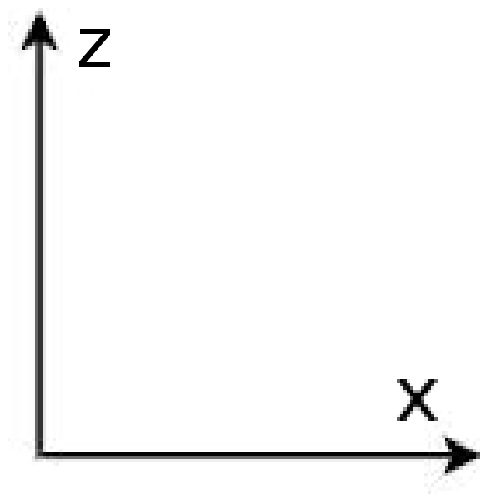}}
\subfigure{\includegraphics[width=0.9\columnwidth]{colorbar_3DKH.eps}}
\caption{(Color Online) Projection of the results presented in Fig.\ref{3DKH} onto the $xz$ plane. Times and colors are the same as Fig.\ref{3DKH}.}
\label{3DKH_xz}
\end{figure}

To set-up a shear velocity in the photosphere, two adjacent vortices rotating in the same direction are produced by applying proper external forces. Therefore, an external force is applied to both sub-domains of $-3\leq x < 0$ by $-3\leq y \leq 3$ by $0\leq z \leq 0.2343$ as well as $0\leq x \leq 3$ by $-3\leq y \leq 3$ by $0\leq z \leq 0.2343$, with the following form:
\begin{equation}
F^x=F_0\sin{(\pi x^*/(3))} \cos{(\pi y/6)},
\end{equation}
\begin{equation}
F^y=F_0\cos{(\pi x^*/(3))} \sin{(\pi y/6)},
\end{equation}
where $F^x$ and $F^y$ are the components of the external force in the $x$ and $y$ directions, respectively, $F_0$ is the magnitude of the applied force, which is set to $0.05$ and $x^*=x$ for the sub-domain $-3\leq x < 0$ and $x^*=x-3$ for the sub-domain $0\leq x \leq 3$. Note that, to limit the magnitude of the velocity, the external force is non-zero only when the maximum velocity in the domain is smaller than $0.3 c$. For the numerical simulation $\sigma=100$, $\Gamma=4/3$, $\delta x/\delta t=2.5\sqrt{2}$, $\tau=1.0$ and $\alpha=0.1$ are considered.

The results of the 3D simulation are shown in Fig.\ref{3DKH}. One can see the initial condition of the magnetic field lines in Fig.\ref{3DKH}(a). The colors on the outer domain boundaries indicate the value of the vorticity, which is zero everywhere in the beginning. After applying the external force, two vortices form, which rotate in the same direction and therefore give rise to a shear velocity in the middle of the $xy$ plane, i.e, the photosphere plane. This shear velocity finally results in a ``cat's eye'' structure in the photosphere (see Fig.\ref{3DKH}(c)) similar to the results of the 2D KH instability with uniform initial density in Fig.\ref{KH_0}. The shear velocity starts to twist the foot of the background magnetic lines and later the upper parts of the background magnetic lines. As a result at some point, the twisted background magnetic lines and the flux rope magnetic lines take opposite directions. This is where the current sheet forms and the reconnection between these two sets of magnetic lines takes place (see Fig.\ref{3DKH}(b)). At late times, Fig. \ref{3DKH}(c), most of the flux rope magnetic lines reconnect with the background magnetic lines, and only a small part of the flux rope can reach the opposite surface. This process can be appreciated more clearly in the 2D projections on the $xz$ plane, which are provided in Fig.\ref{3DKH_xz}. For instance, a starting configuration of the reconnection can be observed in Fig.\ref{3DKH_xz}(b), where one can see that the background magnetic lines reconnect to the flux rope lines and change their topology.

\section {Conclusions}
\label{Conclusions}
We have developed a relativistic MHD lattice Boltzmann model, capable of dealing with problems in the resistive and ideal regimes. The model is based on  the relativistic LB model proposed in Ref.\cite{mohseni2013lattice}, to solve the hydrodynamics equations, and the model proposed in Ref.\cite{mendoza2010three} to solve the Maxwell equations, where several modifications and extensions are implemented to couple the models and to use them in the relativistic MHD context. Thus, a D3Q19 lattice configuration is used for the hydrodynamic part and a D3Q13 lattice for the electromagnetic part.

The numerical method is validated for test simulations in two different regimes, namely propagation of an Alfv\'en wave in the ideal MHD limit (high conductivity), and evolution of a current sheet in a resistive regime (low conductivity). The results are compared with the analytical ones and very good agreement is observed. Additionally, the magnetic reconnection driven by the relativistic KH instability is studied in detail and the effect of different parameters on the reconnection rate is investigated. It is concluded that, while the density ratio has negligible effects on the reconnection rate, an increase in the value of the shear velocity will decrease the reconnection rate. We have also found that, the reconnection rate is proportional to $\sigma^{-\frac{1}{2}}$, which agrees with the scaling law of Sweet-Parker model. Finally, we have presented the results of 3D simulation of the magnetic reconnection in a stellar flare, which is driven by shear velocity in the photosphere. We have shown that due to the shear velocity the reconnection happens between flux rope and background magnetic field lines.

It is worth mentioning that, despite the fact that the model in Ref.~\cite{mohseni2013lattice} is numerically robust at high velocities, the current model becomes numerically unstable when the fluid velocity is high, i.e., in the relativistic supersonic regime. The reason is not fully clear to us at the moment and the issue can be an interesting subject for future extension of the work. Also, not that the relativistic lattice Boltzmann models show nearly
an order of magnitude faster performance than the corresponding hydrodynamic codes \cite{mendoza2010fast,mohseni2013lattice}. Furthermore, the lattice Boltzmann model to solve the Maxwell equations also turns out to be very efficient, almost an order of magnitude faster than Yee's original FDTD method \cite{mendoza2010three}. Thus, we expect our model to have a competitive performance comparing with current models for resistive relativistic MHD.

\begin{acknowledgments}
  We acknowledge financial support from the European Research Council
  (ERC) Advanced Grant 319968-FlowCCS and financial support of the Eidgenössische Technische
  Hochschule Z\"urich (ETHZ) under Grant No. 0611-1.
\end{acknowledgments}
\bibliography{ref_RRMHDLB}

\end{document}